\documentclass[a4paper,12pt]{article}

\usepackage{tikz}
\usetikzlibrary{decorations.pathreplacing,calligraphy}
\usepackage{caption}

\usepackage{verbatim}
\usepackage{amsfonts}
\usepackage{mathrsfs}
\usepackage{amsmath}
\usepackage{amssymb}
\usepackage{framed}
\usepackage[medium]{titlesec}
\usepackage{bm}
\usepackage{cite}
\usepackage[normalem]{ulem}
\usepackage{extarrows}
\usepackage{slashed}
\usepackage{isodateo}
\usepackage{graphicx}
\usepackage{xcolor}
\usepackage[bookmarksnumbered=true,bookmarksopen=true,hyperfootnotes=true]{hyperref}
\hypersetup{colorlinks,%
	linkcolor=[rgb]{0,0.3,0.6}, %
	citecolor=[rgb]{0,0.3,0.6}, %
	urlcolor=[rgb]{0,0.3,0.6}}
\usepackage[hmargin=.7in,vmargin=1.1in]{geometry}
\usepackage{indentfirst}
\usepackage{booktabs}

\usepackage{enumerate}
\usepackage{amsfonts}
\usepackage{mathrsfs}
\usepackage{amsmath}
\usepackage{amssymb}
\usepackage{float}
\usepackage{bm}
\usepackage{slashed}
\usepackage[normalem]{ulem}
\usepackage{extarrows}
\usepackage{slashed}
\usepackage{isodateo}
\usepackage{graphicx}
\usepackage{xcolor}
\usepackage{cancel}

\definecolor{bblue}{rgb}{0,0.2,0.6}

\usepackage{caption}

\usepackage{indentfirst}

\graphicspath{{fig/}}

\usetikzlibrary{arrows,shapes}
\usetikzlibrary{trees}
\usetikzlibrary{matrix,arrows} 				
\usetikzlibrary{positioning}				
\usetikzlibrary{calc,through}				
\usetikzlibrary{decorations.pathreplacing}  
\usepackage{pgffor}							

\usetikzlibrary{decorations.pathmorphing}	
\usetikzlibrary{decorations.markings}
\tikzset{
	photon/.style={decorate, decoration={snake}, draw=red},
	electron/.style={draw=blue, postaction={decorate},
		decoration={markings,mark=at position .55 with {\arrow[draw=blue]{>}}}},
	gluon/.style={decorate, draw=black,
		decoration={coil,amplitude=4pt, segment length=4pt}} ,
	vector/.style={decorate, decoration={snake}, draw},
	provector/.style={decorate, decoration={snake,amplitude=2.5pt}, draw},
	antivector/.style={decorate, decoration={snake,amplitude=-2.5pt}, draw},
	fermion/.style={draw=black, postaction={decorate},
		decoration={markings,mark=at position .55 with {\arrow[draw=black]{>}}}},
	fermionbar/.style={draw=black, postaction={decorate},
		decoration={markings,mark=at position .55 with {\arrow[draw=black]{<}}}},
	fermionnoarrow/.style={draw=black},
	fermionnoarrowsoft/.style={draw=blue},
	scalar/.style={dashed,draw=black, postaction={decorate},
		decoration={markings,mark=at position .55 with {\arrow[draw=black]{>}}}},
	scalarbar/.style={dashed,draw=black, postaction={decorate},
		decoration={markings,mark=at position .55 with {\arrow[draw=black]{<}}}},
	scalarnoarrow/.style={dashed,draw=black},
	scalarnoarrowsoft/.style={dashed,draw=blue},
	electron/.style={draw=black, postaction={decorate},
		decoration={markings,mark=at position .55 with {\arrow[draw=black]{>}}}},
	bigvector/.style={decorate, decoration={snake,amplitude=4pt}, draw},
}

\tikzstyle{block} = [draw, rectangle, 
minimum height=3em, minimum width=6em]
\usetikzlibrary{arrows,patterns}

\newcommand{\email}[1]{\footnote{Email: \href{mailto:#1}{\nolinkurl{#1}}}}

\def\be{\begin{equation}}
	\def\ee{\end{equation}}

\def\bgm{\begin{matrix}}
	\def\edm{\end{matrix}}

\usepackage{mdframed}

\newmdenv[skipabove=0pt,%
skipbelow=5pt,%
leftmargin=0pt,%
rightmargin=0pt,%
innertopmargin=-5pt,%
innerbottommargin=7pt,%
innerleftmargin=2pt,%
innerrightmargin=2pt,%
splittopskip=0pt,%
splitbottomskip=0pt,%
linewidth=0pt,%
nobreak=true]%
{keyeqn2}

\newmdenv[backgroundcolor=gray!15,%
skipabove=0pt,%
skipbelow=5pt,%
leftmargin=0pt,%
rightmargin=0pt,%
innertopmargin=-5pt,%
innerbottommargin=7pt,%
innerleftmargin=2pt,%
innerrightmargin=2pt,%
splittopskip=0pt,%
splitbottomskip=0pt,%
linewidth=0pt,%
nobreak=true]%
{keyeqn}

\usepackage{titlesec}

\begin{document}
	\title{\Large\textbf{Causality and the Interpretation of Quantum Mechanics}\\[2mm]}
	
	\author{
		Kaixun Tu$^{1,\,}$\email{tkx19@tsinghua.org.cn}~~~
            Qing Wang$^{1,\,2,\,}$\email{wangq@mail.tsinghua.edu.cn}\\[5mm]
		\normalsize{${}^{1}\,$\emph{Department of Physics, Tsinghua University, Beijing 100084, China}}\\
            \normalsize{${}^{2}\,$\emph{Center for High Energy Physics, Tsinghua University, Beijing 100084, China}}}
	
	\date{}
	\vspace{20mm}
	\maketitle
	\begin{abstract}
		\vspace{10mm}	
{From the ancient Einstein-Podolsky-Rosen paradox to the recent Sorkin-type impossible measurements problem, the contradictions between relativistic causality, quantum non-locality, and quantum measurement have persisted. Based on quantum field theory, our work provides a framework that harmoniously integrates these three aspects.  This framework consists of causality expressed by reduced density matrices and an interpretation of quantum mechanics that considers quantum mechanics to be complete.
	Specifically, we use reduced density matrices to represent the local information of the quantum state and show that the reduced density matrices cannot evolve superluminally.
	Unlike recent approaches that address causality by introducing new operators to represent detectors, our perspective is that everything—including detectors, the environment, and even humans—is made up of the same fundamental fields. This viewpoint leads us to question the validity of the Schr\"odinger's cat paradox and motivates us to propose an interpretation of quantum mechanics that requires no extra assumptions and remains fully compatible with  relativity.
	}
	\end{abstract}
	\newpage
	\tableofcontents
	
	\newpage

\section{Introduction}
\label{In}

To discuss relativistic causality within the framework of quantum mechanics, we need the concept of locality. For the case of a single-particle state, we can impart quantum state locality by defining position operators \cite{Newton1949}.  However, it leads to superluminal propagation, as succinctly argued in \cite{Kaiser2018}. Further variations and physical implications are discussed in \cite{Hegerfeldt1974,Hegerfeldt1980,Hegerfeldt1985,Rosenstein1987,Barat2003,Wagner2011,Wagner2012,AlHashimi2014,Eckstein2017,Pavsic2018}.
In fact, particles are quantized waves rather than point-like entities.
Due to pair production, we cannot precisely determine the position of a particle \cite{Kaiser2018,Bohr1950,Peres2004}, which means that position operators are not suited for a genuine description of local phenomena in quantum field theory.
Therefore, superluminal propagation induced by the use of position operators is not physically realizable.
Furthermore, our later results indicate that a group of particles initially confined to a certain region will not propagate superluminally in the subsequent evolution.

Compared to superluminal propagation of particles, the ``Fermi two-atom problem" can more accurately reflect the core of causality \cite{Fermi1932}.	
Fermi considered two atoms separated by a distance $R$, with one atom in the ground state and the other in the excited state. If causal influences propagate at the maximum speed $c$, then within the time interval $0<t<R/c$, the excited atom will not have any impact on the ground-state atom through emitted photons.

The Fermi two-atom problem garnered renewed attention when revisited sixty years later by Hegerfeldt \cite{Hegerfeldt1994} (The introduction in \cite{Power1997} is a brief review of earlier work).	
Hegerfeldt fixed the local properties of states with the help
of projection operators, and demonstrated that describing this experiment in terms of transition probabilities leads to results that violate causality.	
Subsequently, Buchholz and Yngvason
used algebraic quantum field theory to analyze the problem and
pointed out that Hegerfeldt's projection operators are not legitimate quantities, and expectation values should be used instead. \cite{Buchholz1994}.
In the modern development of the Fermi two-atom problem, the two atoms were replaced by two Unruh-DeWitt detectors to explore causality \cite{Buscemi2009,Sabin2011,Cliche2010,Jonsson2014,MartinMartinez2015,Dickinson2016,Kolioni2020,Tjoa2022}.
In this context, detectors are not composed of fields used to propagate influences over the distances; instead, one needs to enlarge the Hilbert space to introduce the quantum states of the detectors. Additionally, the initial states are bare states (direct product states), involving the projection operators required by Hegerfeldt \cite{Hegerfeldt1994}, which, however, was precisely the point refuted by Buchholz and Yngvason \cite{Buchholz1994}. 

In the framework of quantum field theory, all matter in the world is composed of the same fundamental fields, which means that we do not need to enlarge the Hilbert space to introduce detectors. 
More specifically, detectors are composed of elementary particles, and elementary particles are excitations of fundamental fields; therefore, detectors are essentially excitations of fundamental fields as well.
Regardless of the number of detectors, the Hilbert space used to describe the entire system should remain the same. 
At the same time, this also implies that we cannot separate detectors from the systems being detected. Even if we consider quantum fields at different spatial points as different field operators, the interactions between fields at adjacent points lead to strong entanglement between different points in the quantum state. This also means that even basic and simple states like the vacuum state cannot be written as the direct product of ``vacuum states" of different regions. 
The characteristic of quantum field theory has been noted by Hegerfeldt (see \cite{Hegerfeldt1994}, last page) and further analyzed by Buchholz and Yngvason \cite{Buchholz1994}. However, subsequent studies and discussions on causality have overlooked this characteristic and simply regarded the initial state as a bare state.
Later, we will point out that the idea that ``all matter in the world is composed of the same fundamental fields" plays an important role in finding an interpretation of quantum mechanics compatible with relativity.



In addition to the two-detector system mentioned above, introducing a third detector will lead to ``Sorkin-type impossible measurements problem" illustrating that the natural generalization of the non-relativistic measurement scheme to relativistic quantum theory leads to superluminal signaling \cite{Sorkin1993,Borsten2021,Papageorgiou2024}.
Traditional quantum mechanics is governed by two laws:  time evolution of quantum states and measurement theory. Relativity requires that these two laws together must ensure that information cannot propagate faster than light. However, non-relativistic measurement theories with traditional state update rule (Lüders’ rule) cannot meet this requirement.
To address this issue, various ``update rules" 
have recently been proposed to describe the state update induced by measurements on the fields or on the detectors  \cite{Borsten2021,Jubb2022,Fewster2020,Bostelmann2021,Ramon2021,PoloGomez2022,Papageorgiou2024}. 
Similar to the Fermi two-atom problem, new operators are introduced to specifically describe the detectors, then the resulting ``update rules" to some extent depend on the model construction.
In fact, the Sorkin-type impossible measurements problem reveals a deeper issue, providing a window for exploring the interpretation of quantum mechanics. Since the physical law behind measurement theory are the interpretation of quantum mechanics \cite{Aharonov2008}, and traditional non-relativistic measurement theory has been shown to be inadequate, it suggests that traditional interpretations might also be flawed. A correct interpretation of quantum mechanics has the potential to provide a measurement theory that satisfies causality.

In addition to the method of introducing detectors as described above, one can also discuss causality by specifying rules for physical operations on quantum states.
Quoting Witten's discussion of the Reeh-Schlieder theorem, he states \cite{Witten2018}, ``If one asks about not mathematical operations in Hilbert space but physical operations that are possible in the real world, then the only physical way that one can modify a quantum state is by perturbing the Hamiltonian by which it evolves, thus bringing about a unitary transformation.”
This classic statement also serves as the starting point for earlier proofs regarding causality \cite{Eberhard1989,Nomoto1992}.
However, the assertion that ``the only physical way that one can modify a quantum state is by `perturbing the Hamiltonian' by which it evolves" has not been robustly demonstrated 
(for a more detailed elaboration on the concept of ``perturbing the Hamiltonian," one can refer to the third hypothesis outlined in \cite{Eberhard1989}).
Especially when humans (or apparatus) interact with the system, it is generally challenging to regard the system as a pure state after the operations due to the entanglement caused by the interaction.
Later, we will demonstrate that, the conclusion that physical operations can always be equivalent to a unitary operator need not rely on the notion of ``perturbing the Hamilton" but can be derived from our results.

In this paper, we adopt a more realistic perspective to investigate causality, where everything (including detectors, environments, and humans) is composed of the same fundamental fields and collectively described by a pure state. This implies that there is no need to enlarge the Hilbert space to introduce detectors, and the Hilbert space of the entire system remains the same regardless of the number of detectors.
Specifically, we employ reduced density matrices to characterize the local information of quantum states.
Dividing the entire space into two regions, denoted as $a$ and $A$, we trace out region $A$ ($a$) to obtain the reduced density matrix which characterizes the information of the quantum state in region $a$ ($A$). 
As shown in Fig.~\ref{1}, region $b$ is the $t$ surface in the future domain of influence of region $a$.	
Consider two quantum states with unequal reduced density matrices of region $a$ and equal reduced density matrices of region $A$.
We will demonstrate that, for local quantum field theory, the reduced density matrices of region $B$ are equal for the two quantum states at time $t$.
Furthermore, since detectors and the systems under measurement are both composed of the same fundamental fields, the ``composite" system formed by a detector and a system under measurement cannot be written in the form of $|\text{System}\rangle|\text{Detector}\rangle$.
Instead, the entire system should be described by the quantum state of quantum field theory, while the system under measurement and the detector can be described by the corresponding reduced density matrices.
However, when two quantum states with the same reduced density matrix in a certain region are superposed, due to spatial entanglement, the superposition state may no longer retains the original reduced density matrix in that region.
This characteristic of quantum field theory implies that there may be no Schrödinger’s cat paradox in quantum field theory and gives rise to an interpretation of quantum mechanics that does not require any additional assumptions (such as the many-worlds hypothesis) and is compatible with relativity.

%
%

This paper is organized as follows.
In Section \ref{lac}, we demonstrate causality using the example of a free scalar field, providing a detailed exposition of our definition of causality. 
Subsequently, we extend the proof of causality to the general case of local quantum field theory. 
In Section \ref{W}, we analyze Witten's statement regarding physical operations when discussing the Reeh-Schlieder theorem.
In Section \ref{inter}, we point out a loophole in the derivation of Schr\"odinger's cat paradox and propose an interpretation of quantum mechanics that supports the completeness of quantum mechanics. 
We also discuss the differences of this interpretation from traditional hidden-variable theories and its compatibility with relativity.
Section \ref{con} concludes the paper with a summary and some discussions, along with several directions for further investigations.
Appendix \ref{Sss} provides the derivation of the field propagator used to compute the time evolution of the density matrix.







\begin{figure}[t]
	\centering	
	\begin{tikzpicture}[line width=0.4pt,scale=0.8,>=stealth]
		\draw(-4,2)--(4,2);
		\draw[->](0,0)--(0,2.6)node[left]{time};
		\draw[->](-4,0)--(4,0)node[below]{space};
		\node at(-0.1,.15){$0$};
		
		\node at(-0.1,2.15){$t$};
		
		\node at(-2.5,0.5){$A$};
		\node at(2.5,0.5){$A$};
		\node at(-3,2.4){$B$};
		\node at(3,2.4){$B$};
		\node at(0,-0.35){$a$};
		\node at(-0.7,1.6){$b$};
		
		\draw(-1,0)--(-2,2);
		\draw(1,0)--(2,2);
		\draw [decorate, 
		decoration = {calligraphic brace, 
			raise=0pt, 
			aspect=0.5, 
			amplitude=6pt 
		}] (-4,0) --  (-1,0);
		\draw [decorate, 
		decoration = {calligraphic brace, 
			raise=0pt, 
			aspect=0.5, 
			amplitude=6pt 
		}] (1,0) --  (4,0);
		\draw [decorate, 
		decoration = {calligraphic brace, 
			raise=0pt, 
			aspect=0.5, 
			amplitude=6pt 
		}] (1,0) --  (-1,0);
		\draw [decorate, 
		decoration = {calligraphic brace, 
			raise=0pt, 
			aspect=0.5, 
			amplitude=6pt 
		}] (-4,2) --  (-2,2);
		\draw [decorate, 
		decoration = {calligraphic brace, 
			raise=0pt, 
			aspect=0.5, 
			amplitude=6pt 
		}] (2,2) --  (4,2);
		\draw [decorate, 
		decoration = {calligraphic brace, 
			raise=0pt, 
			aspect=0.68, 
			amplitude=6pt 
		}] (2,2) --  (-2,2);
		
		\node at(0,-1){$\text{tr}_a\left(|\psi_1\rangle \langle\psi_1|\right) = \text{tr}_a\left(|\psi_2\rangle \langle\psi_2|\right)$};
		
		\node at(0,3.2){$\text{tr}_b\left(|\psi_1;t\rangle \langle\psi_1;t|\right) = \text{tr}_b\left(|\psi_2;t\rangle \langle\psi_2;t|\right)$};
		
		\node at(0,3.2){$?$};
		
	\end{tikzpicture}
	\captionsetup{justification=centering}
	\captionsetup{labelsep=space}
	\caption{\label{1}   The relationships among regions $A$, $a$, $B$, and $b$}
	
\end{figure}

\section{Localization and Causality}
\label{lac}

Ref. \cite{Wagner2011} argues that among various free quantum field theories, the simplest one—the free real scalar field theory (referred to as the ``relativistic Schr\"odinger system" in Ref. \cite{Wagner2011})—actually violates causality.
We do not agree with this conclusion. Therefore, in order to clearly demonstrate our definition of causality, we first use the free real scalar field as an example to illustrate and prove causality. Subsequently, we will prove causality for general local quantum field theories.

\subsection{Free Field Theory as An Example}
\label{Ff}

Consider a free field theory described by the Hamiltonian
$\hat H 
=\int \mathrm d^3x
\Big[\frac{1}{2}\hat\pi^2(\boldsymbol{x})
+\frac{1}{2}\Big(\nabla\hat\phi(\boldsymbol{x})\Big)^2
+\frac{1}{2}m^2\hat\phi^2(\boldsymbol{x})\Big]$, 
and a quantum state $|\psi;t\rangle=\text{e}^{-i\hat H t}|\psi\rangle$ evolving from an initial state $|\psi\rangle$ after a time $t$.
In fact, the expression of $H$ in terms of creation and annihilation operators is precisely the same as $\hat{H}_{rS}$ in equation (3.3a) of Ref. \cite{Wagner2011}, which describes the relativistic Schr\"odinger system.
To highlight the local information of the quantum state, we use a representation based on the eigenstates $|\phi\rangle$ of the field operator $\hat\phi$, rather than the traditional Fock representation.
These states are normalized according to $\langle\phi|\phi'\rangle=\prod\limits_{\boldsymbol{x}} \delta(\phi(\boldsymbol{x})-\phi'(\boldsymbol{x}))$. Sometimes, we interchange the symbol $\phi$ with $\varphi$. 
Utilizing the field propagator derived in Appendix \ref{Sss}, the density matrix $\hat\rho(t)=|\psi;t\rangle \langle\psi;t|$ can be expressed as
\begin{equation}
	\begin{split}
		\rho(\phi,\phi';t)
		&=\langle\phi|\psi;t \rangle \langle\psi;t|\phi'\rangle\\
		&=|N(t)|^2  
		\int\mathcal{D}\varphi   \mathcal{D}\varphi'\;
		\rho(\varphi,\varphi')
		K(\phi,\phi',\varphi,\varphi';t)
		\; ,
	\end{split}
\end{equation}
where
\begin{eqnarray}
	\rho(\varphi,\varphi')&=&\langle\varphi|\psi\rangle \langle\psi|\varphi'\rangle\;,
	\\	
	N(t)&=&\mathcal N \text{e}^{-\frac{1}{2}\int \mathrm d^3x\int \mathrm dt\; G(\boldsymbol{0};t)}\;,
	\\
	K(\phi,\phi',\varphi,\varphi';t)&=&\text{e}^{iS(\phi,\varphi;t)-iS(\phi',\varphi';t)}
	\label{K}
	\; ,
\end{eqnarray}
\begin{equation}\label{s}
	\begin{split}
		S(\phi,\varphi;t)
		=&\frac{1}{2}\int \mathrm d^3x \mathrm d^3y\; G(\boldsymbol{x}-\boldsymbol{y};t)\Big[\phi(\boldsymbol{x})\phi(\boldsymbol{y})
		+\varphi(\boldsymbol{x})\varphi(\boldsymbol{y})\Big]\\
		&-\int \mathrm d^3x \mathrm d^3y\; g(\boldsymbol{x}-\boldsymbol{y};t)\phi(\boldsymbol{x})\varphi(\boldsymbol{y}) \;
		,
	\end{split}
\end{equation}and
\begin{equation}
	\begin{split}
		\label{gG}
		G(\boldsymbol{x}-\boldsymbol{y};t)
		&=\int \frac{\mathrm d^3p}{(2\pi)^3} 	 \frac{p^0 \cos(p^0t)}{\sin(p^0t)}
		\text{e}^{i\boldsymbol{p}\cdot(\boldsymbol{x}-\boldsymbol{y}) }
		\; ,   \\
		g(\boldsymbol{x}-\boldsymbol{y};t)          
		&=\int \frac{\mathrm d^3p}{(2\pi)^3} 	\frac{p^0}{\sin(p^0t)}
		\text{e}^{i\boldsymbol{p}\cdot(\boldsymbol{x}-\boldsymbol{y}) }
		\; .
	\end{split}
\end{equation}

Dividing the space into two parts, denoted as $B$ and $b$, with the values of $\phi$ on the two regions denoted as $\phi_B$ and $\phi_b$, the reduced density matrix of the region $B$ in Fig.~\ref{1} is given by
\begin{equation}
	\begin{split}
		\label{B}
		\int \mathcal{D}\phi_b\; 
		\rho(\phi,\phi';t)\bigg|_{\phi_b=\phi'_b}
		=|N(t)|^2  
		\int\mathcal{D}\varphi   \mathcal{D}\varphi'
		\rho(\varphi,\varphi')
		\int \mathcal{D}\phi_b\;K(\phi,\phi',\varphi,\varphi';t)
		\bigg|_{\phi_b=\phi'_b}
		\; .
	\end{split}
\end{equation}
Further utilization of Eq. \eqref{K} and Eq. \eqref{s} yields the final integral of Eq. \eqref{B} as

\begin{equation}
	\begin{split}
		\label{s-s}
		&\int \mathcal{D}\phi_b\;K(\phi,\phi',\varphi,\varphi';t)
		\bigg|_{\phi_b=\phi'_b}\\
		&=
		\exp\Bigg\{\frac{i}{2}\int_B \mathrm d^3x \int_B \mathrm d^3y \;
		G(\boldsymbol{x}-\boldsymbol{y};t)
		\Big[\phi_B(\boldsymbol{x})\phi_B(\boldsymbol{y})
		-\phi'_B(\boldsymbol{x})\phi'_B(\boldsymbol{y})\Big]\\
		&\qquad\qquad+
		\frac{i}{2}\int \mathrm d^3x \int \mathrm d^3y \;
		G(\boldsymbol{x}-\boldsymbol{y};t)
		\Big[\varphi(\boldsymbol{x})\varphi(\boldsymbol{y})
		-\varphi'(\boldsymbol{x})\varphi'(\boldsymbol{y})\Big]\\
		&\qquad\qquad-
		i\int_B \mathrm d^3x \int \mathrm d^3y\;
		g(\boldsymbol{x}-\boldsymbol{y};t)
		\Big[\phi_B(\boldsymbol{x})\varphi(\boldsymbol{y})
		-\phi'_B(\boldsymbol{x})\varphi'(\boldsymbol{y})\Big]\Bigg\}
		\\
		&\quad\times\prod\limits_{\boldsymbol{x}\in b}
		\delta\Bigg[  \int_B \mathrm d^3y\;	G(\boldsymbol{x}-\boldsymbol{y};t)
		\left[\phi_B(\boldsymbol{y})-\phi'_B(\boldsymbol{y})\right]
		-
		\int \mathrm d^3y \; g(\boldsymbol{x}-\boldsymbol{y};t)
		\left[\varphi(\boldsymbol{y})-\varphi'(\boldsymbol{y})\right]
		\Bigg] .
	\end{split}
\end{equation}

According to Eq. \eqref{gG}, the inverse of the function $g(\boldsymbol{x}-\boldsymbol{y};t)$ can be determined as
\begin{equation}\label{-1}
	\begin{split}
		g_{-1}(\boldsymbol{x}-\boldsymbol{y};t)
		&=\int \frac{d^3p}{(2\pi)^3} 	\frac{\sin(p^0t)}{p^0}
		e^{i\boldsymbol{p}\cdot(\boldsymbol{x}-\boldsymbol{y}) }
		\; .
	\end{split}
\end{equation}
It is easy to verify that the functions $g(\boldsymbol{x}-\boldsymbol{y};t)$ and $g_{-1}(\boldsymbol{x}-\boldsymbol{y};t)$ satisfy
\begin{equation}\label{gg}
	\begin{split}
		\int \mathrm d^3y\;
		g_{-1}(\boldsymbol{x}-\boldsymbol{y};t)
		g(\boldsymbol{y}-\boldsymbol{z};t)
		=\delta(\boldsymbol{x}-\boldsymbol{z})
		\; .
	\end{split}
\end{equation}

Based on Eq. \eqref{gG} and Eq. \eqref{-1}, we can also obtain the following useful formulas:
\begin{equation}\label{gG1}
	\begin{split}
		\int \mathrm d^3y\;g_{-1}(\boldsymbol{x}-\boldsymbol{y};t)
		G(\boldsymbol{y}-\boldsymbol{z};t)
		=\frac{\partial}{\partial t}
		g_{-1}(\boldsymbol{x}-\boldsymbol{z};t)\; ,
	\end{split}
\end{equation}
\begin{equation}\label{gG2}
	\begin{split}
		\int \mathrm d^3z\;
		\frac{\partial}{\partial t}g_{-1}(\boldsymbol{x}-\boldsymbol{z};t)
		G(\boldsymbol{z}-\boldsymbol{y};t)
		= g(\boldsymbol{x}-\boldsymbol{y};t)
		+\frac{\partial^2}{\partial t^2} g_{-1}(\boldsymbol{x}-\boldsymbol{y};t)\; .
	\end{split}
\end{equation}


Based on Eq. \eqref{-1}, we also note an important equality: $g_{-1}(\boldsymbol{x}-\boldsymbol{y};t)=\left[\hat\phi(\boldsymbol{x},t),\hat\phi(\boldsymbol{y},0)\right]$, where $\hat\phi(\boldsymbol{x},t)$ is the field operator in the Heisenberg picture. Therefore, it is evident that
\begin{equation}
	\begin{split}
		\label{sl}
		g_{-1}(\boldsymbol{x};t)=0 \;,\quad\boldsymbol{x}^2-t^2>0\; .
	\end{split}
\end{equation}
As shown in Fig.~\ref{1}, the region $a$ is the $t=0$ surface in the past domain of dependence of region $b$.	
Combining Fig.~\ref{1} with Eq. \eqref{sl}, we can rewrite Eq. \eqref{gg}, Eq. \eqref{gG1}, and Eq. \eqref{gG2} as follows:
\begin{equation}\label{gg0}
	\begin{split}
		\int_b \mathrm d^3y\;
		g_{-1}(\boldsymbol{x}-\boldsymbol{y};t)
		g(\boldsymbol{y}-\boldsymbol{z};t)
		=\delta(\boldsymbol{x}-\boldsymbol{z})
		,\;  \boldsymbol{x}\in a ,
	\end{split}
\end{equation}
\begin{equation}\label{gG01}
	\begin{split}
		\int_b \mathrm d^3y\;g_{-1}(\boldsymbol{x}-\boldsymbol{y};t)
		G(\boldsymbol{y}-\boldsymbol{z};t)
		=\frac{\partial}{\partial t}
		g_{-1}(\boldsymbol{x}-\boldsymbol{z};t)
		,\;  \boldsymbol{x}\in a ,
	\end{split}
\end{equation}
\begin{equation}\label{gG02}
	\begin{split}
		\int_b \mathrm d^3z\;
		\frac{\partial}{\partial t}g_{-1}(\boldsymbol{x}-\boldsymbol{z};t)
		G(\boldsymbol{z}-\boldsymbol{y};t)
		= g(\boldsymbol{x}-\boldsymbol{y};t)
		+\frac{\partial^2}{\partial t^2} g_{-1}(\boldsymbol{x}-\boldsymbol{y};t)
		,\; \boldsymbol{x}\in a .
	\end{split}
\end{equation}

Let's focus back on Eq. \eqref{s-s}. The $\delta$ function in Eq. \eqref{s-s} indicates that
\begin{equation}\label{d1}
	\begin{split}
		\int \mathrm d^3y \; g(\boldsymbol{x}-\boldsymbol{y};t)
		\left[\varphi(\boldsymbol{y})-\varphi'(\boldsymbol{y})\right]
		= \int_B \mathrm d^3y\;	G(\boldsymbol{x}-\boldsymbol{y};t)
		\left[\phi_B(\boldsymbol{y})-\phi'_B(\boldsymbol{y})\right]
		,\; \boldsymbol{x}\in b.
	\end{split}
\end{equation}
Based on Eq. \eqref{gg0}, Eq. \eqref{d1}, Eq. \eqref{gG01}, and Eq. \eqref{sl},  in region $a$ we can obtain
\begin{equation}\label{=0}
	\begin{split}
		&\varphi_a(\boldsymbol{x})-\varphi'_a(\boldsymbol{x})\\
		&=	\int \mathrm d^3y\; 
		\delta(\boldsymbol{x}-\boldsymbol{y})
		\left[\varphi_a(\boldsymbol{y})-\varphi'_a(\boldsymbol{y})\right]\\
		&=	\int \mathrm d^3y
		\int_b \mathrm d^3z\;
		g_{-1}(\boldsymbol{x}-\boldsymbol{z};t)
		g(\boldsymbol{z}-\boldsymbol{y};t)
		\left[\varphi_a(\boldsymbol{y})-\varphi'_a(\boldsymbol{y})\right]\\
		&=	\int_B \mathrm d^3y
		\int_b \mathrm d^3z\;
		g_{-1}(\boldsymbol{x}-\boldsymbol{z};t)
		G(\boldsymbol{z}-\boldsymbol{y};t)
		\left[\phi_B(\boldsymbol{y})-\phi'_B(\boldsymbol{y})\right]\\
		&=	\int_B \mathrm d^3y
		\int_b \mathrm d^3z\;
		g_{-1}(\boldsymbol{x}-\boldsymbol{z};t)
		G(\boldsymbol{z}-\boldsymbol{y};t)
		\left[\phi_B(\boldsymbol{y})-\phi'_B(\boldsymbol{y})\right]\\
		&=0  \;,\qquad\qquad\qquad\qquad    \boldsymbol{x}\in a,
	\end{split}
\end{equation}
where, the second equality employs Eq. \eqref{gg0}, the third equality employs Eq. \eqref{d1}, the fourth equality employs Eq. \eqref{gG01}, and the final equality employs Eq. \eqref{sl}.
Then, according to Eq. \eqref{=0}, the $\delta$ function in \eqref{s-s} can be expressed as
\begin{equation}\label{delta}
	\begin{split}		
		\propto
		&\prod\limits_{\boldsymbol{x}\in a}
		\delta
		\Big(\varphi_a(\boldsymbol{x})-\varphi'_a(\boldsymbol{x})\Big)\\
		&\times
		\prod\limits_{\boldsymbol{x}\in b}
		\delta
		\Bigg(  \int_B \mathrm d^3y\;	G(\boldsymbol{x}-\boldsymbol{y};t)
		\left[\phi_B(\boldsymbol{y})-\phi'_B(\boldsymbol{y})\right]\\
		&\qquad\qquad\quad-
		\int_A \mathrm d^3y\; g(\boldsymbol{x}-\boldsymbol{y};t)
		\left[\varphi_A(\boldsymbol{y})-\varphi'_A(\boldsymbol{y})\right]
		\Bigg)\; .
	\end{split}
\end{equation}
The proportionality coefficient in Eq. \eqref{delta} is independent of $\varphi_a$ and $\varphi'_a$.
Substituting Eq. \eqref{delta} into Eq. \eqref{s-s}, we obtain

\begin{equation}\label{s2}
	\begin{split}
		&\int \mathcal{D}\phi_b\;K(\phi,\phi',\varphi,\varphi';t)
		\bigg|_{\phi_b=\phi'_b}\\
		&\propto 
		\exp\Bigg\{
		i\int_a \mathrm d^3x\; \varphi_a(\boldsymbol{x})
		\Bigg[\int_A \mathrm d^3y \;
		G(\boldsymbol{x}-\boldsymbol{y};t)
		\left[\varphi_A(\boldsymbol{y})-\varphi'_A(\boldsymbol{y})\right] \\
		&~~~~~~~~~~~~~~~~~~~~~~~~~~~~~~~~~~~~
		-\int_B \mathrm d^3y\; g(\boldsymbol{x}-\boldsymbol{y};t)
		\left[\phi_B(\boldsymbol{y})-\phi'_B(\boldsymbol{y})\right]\Bigg] \Bigg\}
		\\
		&\times\exp\Bigg\{
		\frac{i}{2}\int_B \mathrm d^3x \int_B \mathrm d^3y \;
		G(\boldsymbol{x}-\boldsymbol{y};t)
		\Big[\phi_B(\boldsymbol{x})\phi_B(\boldsymbol{y})
		-\phi'_B(\boldsymbol{x})\phi'_B(\boldsymbol{y})\Big]
		\\&
		\qquad\qquad+\frac{i}{2}\int_A \mathrm d^3x \int_A \mathrm d^3y \;
		G(\boldsymbol{x}-\boldsymbol{y};t)
		\Big[\varphi_A(\boldsymbol{x})\varphi_A(\boldsymbol{y})
		-\varphi'_A(\boldsymbol{x})\varphi'_A(\boldsymbol{y})\Big]\\
		&\qquad\qquad\quad-i\int_A \mathrm d^3x \int_B \mathrm d^3y\; g(\boldsymbol{x}-\boldsymbol{y};t)
		\Big[\varphi_A(\boldsymbol{x})\phi_B(\boldsymbol{y})
		-\varphi'_A(\boldsymbol{x})\phi'_B(\boldsymbol{y})\Big]\;
		\Bigg\}
		\\
		&\times	 
		\prod\limits_{\boldsymbol{x}\in b}\delta
		\Bigg(  \int_B \mathrm d^3y\;	G(\boldsymbol{x}-\boldsymbol{y};t)
		\left[\phi_B(\boldsymbol{y})-\phi'_B(\boldsymbol{y})\right]
		-
		\int_A \mathrm d^3y\; g(\boldsymbol{x}-\boldsymbol{y};t)
		\left[\varphi_A(\boldsymbol{y})-\varphi'_A(\boldsymbol{y})\right]
		\Bigg)
		\\
		&\times \prod\limits_{\boldsymbol{x}\in a}
		\delta
		\Big(\varphi_a(\boldsymbol{x})-\varphi'_a(\boldsymbol{x})\Big)	\;.
	\end{split}
\end{equation}

Next, we proceed to prove that the first two lines (the first exponential term) of Eq. \eqref{s2} equals $1$. Note that the first $\delta$ function in Eq. \eqref{s2} implies
\begin{equation}\label{d2}
	\begin{split}
		\int_B \mathrm d^3y\;	G(\boldsymbol{x}-\boldsymbol{y};t)
		\left[\phi_B(\boldsymbol{y})-\phi'_B(\boldsymbol{y})\right]
		=
		\int_A \mathrm d^3y\; g(\boldsymbol{x}-\boldsymbol{y};t)
		\left[\varphi_A(\boldsymbol{y})-\varphi'_A(\boldsymbol{y})\right]
		,\; \boldsymbol{x}\in b.
	\end{split}
\end{equation}
According to Eq. \eqref{gg0}, Eq. \eqref{gG01}, Eq. \eqref{sl}, and Eq. \eqref{d2}, we can derive the following expression for $\boldsymbol{x}\in a$:
\begin{equation}
	\begin{split}
		\label{d3}
		&\int_A \mathrm d^3y \;
		G(\boldsymbol{x}-\boldsymbol{y};t)
		\Big[\varphi_A(\boldsymbol{y})-\varphi'_A(\boldsymbol{y})\Big]\\
		&=\int_A \mathrm d^3y \int \mathrm d^3s\;
		G(\boldsymbol{x}-\boldsymbol{s};t)
		\delta(\boldsymbol{s}-\boldsymbol{y})
		\Big[\varphi_A(\boldsymbol{y})-\varphi'_A(\boldsymbol{y})\Big]\\
		&=\int \mathrm d^3z\int \mathrm d^3s\;
		G(\boldsymbol{x}-\boldsymbol{s};t)
		g_{-1}(\boldsymbol{s}-\boldsymbol{z};t)
		\int_A \mathrm d^3y\; g(\boldsymbol{z}-\boldsymbol{y};t)
		\Big[\varphi_A(\boldsymbol{y})-\varphi'_A(\boldsymbol{y})\Big]\\ 
		&=\int \mathrm d^3z\;		
		\frac{\partial}{\partial t}	g_{-1}(\boldsymbol{x}-\boldsymbol{z};t)
		\int_A \mathrm d^3y\; g(\boldsymbol{z}-\boldsymbol{y};t)
		\Big[\varphi_A(\boldsymbol{y})-\varphi'_A(\boldsymbol{y})\Big]\\  
		&=\int_b \mathrm d^3z\;		
		\frac{\partial}{\partial t}	g_{-1}(\boldsymbol{x}-\boldsymbol{z};t)
		\int_A \mathrm d^3y\; g(\boldsymbol{z}-\boldsymbol{y};t)
		\Big[\varphi_A(\boldsymbol{y})-\varphi'_A(\boldsymbol{y})\Big]\\ 
		&=\int_b \mathrm d^3z\;		
		\frac{\partial}{\partial t}	g_{-1}(\boldsymbol{x}-\boldsymbol{z};t)
		\int_B \mathrm d^3y\;	G(\boldsymbol{x}-\boldsymbol{y};t)
		\left[\phi_B(\boldsymbol{y})-\phi'_B(\boldsymbol{y})\right], 
	\end{split}
\end{equation}
where, the second equality uses Eq. \eqref{gg0}, the third equality uses Eq. \eqref{gG01}, the fourth equality uses Eq. \eqref{sl}, and the final equality uses Eq. \eqref{d2}.
Furthermore, utilizing Eq. \eqref{gG02}, Eq. \eqref{d3} can be written as
\begin{equation}
	\begin{split}
		\label{delta2}
		&\int_A \mathrm d^3y \;
		G(\boldsymbol{x}-\boldsymbol{y};t)
		\Big[\varphi_A(\boldsymbol{y})-\varphi'_A(\boldsymbol{y})\Big] 
		-\int_B \mathrm d^3y\; g(\boldsymbol{x}-\boldsymbol{y};t)
		\Big[\phi_B(\boldsymbol{y})-\phi'_B(\boldsymbol{y})\Big]\\
		&=
		\int_B \mathrm d^3y\; 
		\frac{\partial^2}{\partial t^2} g_{-1}(\boldsymbol{x}-\boldsymbol{y};t)
		\Big[\phi_B(\boldsymbol{y})-\phi'_B(\boldsymbol{y})\Big]\\
		&=0\;,
	\end{split}
\end{equation}
where $\boldsymbol{x}\in a$ and the last equality utilizes Eq. \eqref{sl}.
From Eq. \eqref{delta2}, it follows that the first two lines (the first exponential term) of Eq.\eqref{s2} equals $1$.
Therefore, utilizing Eq. \eqref{delta2} and Eq. \eqref{s2}, Eq. \eqref{B} can be expressed in the following form:
\begin{equation}
	\begin{split}
		&\int \mathcal{D}\phi_b\; 
		\rho(\phi,\phi';t)\bigg|_{\phi_b=\phi'_b}
		=
		\int\mathcal{D}\varphi_A   \mathcal{D}\varphi'_A
		F(\phi_B,\phi'_B,\varphi_A,\varphi'_A;t)
		\int \mathcal{D}\varphi_a \;
		\rho(\varphi,\varphi')\bigg|_{\varphi_a=\varphi'_a} .
	\end{split}
\end{equation}
This indicates that the reduced density matrix of region $B$ at time $t$ in Fig.~\ref{1} is only determined by the reduced density matrix of region $A$ at time $t=0$, and \eqref{sl} is the key factor leading to this causality.

If the quantum state $|\psi\rangle$ describes a group of particles (which may not be an eigenstate of the particle number) confined in region $a$ at $t=0$,
the quantum state in region $A$ is indistinguishable from the vacuum $|\Omega\rangle$, i.e., $\text{tr}_a\left(|\psi\rangle \langle\psi|\right) = \text{tr}_a\left(|\Omega\rangle \langle\Omega|\right)$.
According to the established causality, for $|\psi;t\rangle=\text{e}^{-i\hat H t}|\psi\rangle$, we have $\text{tr}_b\left(|\psi;t\rangle \langle\psi;t|\right) = \text{tr}_b\left(|\Omega\rangle \langle\Omega|\right)$, which means that at time $t$ the quantum state in region $B$ remains indistinguishable from the vacuum state, indicating that the propagation speed of particles does not exceed the speed of light.

\subsection{General Cases}
\label{Gc}

Ref. \cite{Eberhard1989} demonstrated that: 
For a quantum field theory that can be expressed in the form $\hat H=\int \mathrm d^3x\; \hat h(\boldsymbol{x},t)$,
the reduced density matrix of spacelike-separated regions will not be affected by the action of a human, if ``the action of a human on quantities defined at some point of coordinates $x$ and $t$ results only in changes $\Delta h(x,t)$ of the Hamiltonian density operator $h(x,t)$ defined at the same point" (refer to the third hypothesis in \cite{Eberhard1989}).
Ignoring the physical aspects and focusing solely on the mathematical expression,  we can rephrase the mathematical conclusion in the aforementioned demonstration of Ref. \cite{Eberhard1989} using the language of our paper as follows: 
If there exists a unitary operator $\hat U_a$ supported in region $a$ such that $|\psi_2\rangle=\hat U_a|\psi_1\rangle$ and the relationship between regions $a$ and $b$ is as shown in Fig.~\ref{1}, then we have $\text{tr}_b\left(|\psi_1;t\rangle \langle\psi_1;t|\right) = \text{tr}_b\left(|\psi_2;t\rangle \langle\psi_2;t|\right)$.

Next, we will prove that if two quantum states $|\psi_1\rangle$ and $|\psi_2\rangle$ satisfy
$\text{tr}_a\left(|\psi_1\rangle \langle\psi_1|\right)
=\text{tr}_a\left(|\psi_2\rangle \langle\psi_2|\right)$, then there must exist a unitary operator $\hat U_a$ supported in region $a$ such that $|\psi_2\rangle=\hat U_a|\psi_1\rangle$.


Denoting the basis of the Hilbert space in regions $A$ and $a$ as $|A_n\rangle$ and $|a_m\rangle$ respectively, the quantum states $|\psi_2\rangle$ and $|\psi_1\rangle$ can be expressed as
\begin{equation}\label{aA}
	\begin{split}
		|\psi_i\rangle &=\sum_{m=1}^{M}\sum_{n=1}^{N}f_i(m,n)|a_m\rangle|A_n\rangle
		\; ,\quad i=1,2
	\end{split}
\end{equation}
Let $\boldsymbol{f}_i$ denote an $M\times N$ matrix where $f_i(m, n)$ is the matrix element at the $m$-th row and $n$-th column,			
then the condition $\text{tr}_a\left(|\psi_1\rangle \langle\psi_1|\right)
=\text{tr}_a\left(|\psi_2\rangle \langle\psi_2|\right)$  can be expressed as	
\begin{equation}
	\begin{split}
		\label{ff}
		\boldsymbol{f}^\dagger_1 \boldsymbol{f}_1 =\boldsymbol{f}^\dagger_2 \boldsymbol{f}_2\;.
	\end{split}
\end{equation}

Suppose the rank of the $\boldsymbol{f}_2$ is $k=M$. For the case where $N=M$, $\boldsymbol{f}_2$ has an inverse and then we have $\boldsymbol{f}_2=(\boldsymbol{f}^\dagger_2)^{-1} \boldsymbol{f}^\dagger_1 \boldsymbol{f}_1$ from Eq. \eqref{ff}. Consequently, a unitary matrix $\boldsymbol{U}\equiv(\boldsymbol{f}^\dagger_2)^{-1} \boldsymbol{f}^\dagger_1$ can be defined such that $\boldsymbol{f}_2=\boldsymbol{U} \boldsymbol{f}_1$.	
For the case where $N>M$, it is possible to select $M$ linearly independent column vectors from $\boldsymbol{f}_2$.
Then we can form an $M\times M$ square matrix $\boldsymbol{F}_2$ using these $M$ linearly independent vectors, and it is evident that $\boldsymbol{F}_2$ is an invertible matrix.
Based on the positions of the $M$ linearly independent vectors in $\boldsymbol{f}_2$, we can also correspondingly select column vectors from $\boldsymbol{f}_1$ to form another $M\times M$ square matrix $\boldsymbol{F}_1$.  
According to the definitions of $\boldsymbol{F}_1$ and $\boldsymbol{F}_2$, Eq. \eqref{ff} immediately yields $\boldsymbol{F}^\dagger_1 \boldsymbol{f}_1 = \boldsymbol{F}^\dagger_2 \boldsymbol{f}_2$.	
Therefore, a unitary matrix $\boldsymbol{U}\equiv(\boldsymbol{F}^\dagger_2)^{-1} \boldsymbol{F}^\dagger_1$ can be defined such that $\boldsymbol{f}_2 = \boldsymbol{U} \boldsymbol{f}_1$.

Suppose the rank of the $\boldsymbol{f}_2$ is $k<M$.
According to basic linear algebra knowledge, we know that
$\text{rank}\boldsymbol{f}_1=\text{rank}(\boldsymbol{f}^\dagger_1\boldsymbol{f}_1)$
and
$\text{rank}\boldsymbol{f}_2=\text{rank}(\boldsymbol{f}^\dagger_2\boldsymbol{f}_2)$.
Thus, Eq. \eqref{ff} indicates that the rank of $\boldsymbol{f}_1$ is
$\text{rank}\boldsymbol{f}_1=\text{rank}\boldsymbol{f}_2=k$.
It is possible to add $M-k$ normalized and mutually orthogonal column vectors to $\boldsymbol{f}_1$ ($\boldsymbol{f}_2$), such that these newly added vectors are orthogonal to the original ones in $\boldsymbol{f}_1$ ($\boldsymbol{f}_2$). After adding these vectors, $\boldsymbol{f}_1$ and $\boldsymbol{f}_2$ become extended matrices with rank $M$, denoted as $\boldsymbol{f}'_1$ and $\boldsymbol{f}'_2$.
Due to the orthogonality and normalization of the new column vectors, as well as their orthogonality with the old column vectors, it is evident that \eqref{ff} can be extended to $\boldsymbol{f}'^\dagger_1 \boldsymbol{f}'_1 = \boldsymbol{f}'^\dagger_2 \boldsymbol{f}'_2 $. Consequently, according to the proof in the preceding paragraph, there exists a unitary matrix $\boldsymbol{U}$ such that $\boldsymbol{f}'_2 = \boldsymbol{U} \boldsymbol{f}'_1$. Focusing only on the transformation of the old column vectors, we immediately have $\boldsymbol{f}_2 = \boldsymbol{U} \boldsymbol{f}_1$.


In summary, we can always find a unitary matrix $\boldsymbol{U}$ such that $\boldsymbol{f}_2 = \boldsymbol{U} \boldsymbol{f}_1$.
%
Let $U(m,m')$ be the matrix element of $\boldsymbol{U}$ at the $m$-th row and $m'$-th column. Using Eq. \eqref{aA} and $\boldsymbol{f}_2 = \boldsymbol{U} \boldsymbol{f}_1$, we obtain
\begin{equation}
	\begin{split}\label{U}
		|\psi_2\rangle
		&=\sum_{m=1}^{M}\sum_{n=1}^{N}f_2(m,n)|a_m\rangle|A_n\rangle\\
		&=\sum_{m=1}^{M}\sum_{n=1}^{N}\sum_{m'=1}^{M}U(m,m')f_1(m',n)|a_m\rangle|A_n\rangle\\
		&=\left(\sum_{m=1}^{M}\sum_{m'=1}^{M} U(m,m') |a_m\rangle \langle a_{m'}|\right)|\psi_1\rangle\;.
	\end{split}
\end{equation}	
Eq. \eqref{U} indicates that there exist a unitary operator $\hat U_a$ supported in region $a$ such that $|\psi_2\rangle=\hat U_a|\psi_1\rangle$.
As mentioned at the beginning of this section, combining this with the results from Ref. \cite{Eberhard1989}, we can establish causality: 
In Fig. \ref{1}, if two quantum states $|\psi_1\rangle$ and $|\psi_2\rangle$ initially satisfy $\text{tr}_a\left(|\psi_1\rangle \langle\psi_1|\right) =\text{tr}_a\left(|\psi_2\rangle \langle\psi_2|\right)$, then after time $t$, the two quantum states satisfy $\text{tr}_b\left(|\psi_1;t\rangle \langle\psi_1;t|\right) = \text{tr}_b\left(|\psi_2;t\rangle \langle\psi_2;t|\right)$.

\section{Physical Operations}
\label{W}

Let's analyze Witten's statement \cite{Witten2018} quoted in Section \ref{In}: ``If one asks about not mathematical operations in Hilbert space but physical operations that are possible in the real world, then the only physical way that one can modify a quantum state is by perturbing the Hamiltonian by which it evolves, thus bringing about a unitary transformation.” 

Consider two quantum states: the state $|\psi_1\rangle$ describes only the system being operated on, whereas the other state $|\psi_2\rangle$ includes not only that same system but also an apparatus (or a human) localized in region $a$.
At time $t=0$, the apparatus in region $a$ has not yet performed any operation on the system, so the two quantum states are identical outside region $a$, specifically satisfying $\text{tr}_a\left(|\psi_1\rangle \langle\psi_1|\right) = \text{tr}_a\left(|\psi_2\rangle \langle\psi_2|\right)$.
The apparatus complete the operation after $\Delta t$, and the two quantum states evolve into $|\psi_1;\Delta t\rangle$ and $|\psi_2;\Delta t\rangle$, respectively. Combining with Fig.~\ref{1} (where $t=\Delta t$), causality ensures that $\text{tr}_b\left(|\psi_1;\Delta t\rangle \langle\psi_1;\Delta t|\right) = \text{tr}_b\left(|\psi_2;\Delta t\rangle \langle\psi_2;\Delta t|\right)$.
Based on the conclusion in Section \ref{Gc}, this indicate the existence of a unitary operator $\hat U_b$ supported in region $b$ such that $|\psi_2;\Delta t\rangle =\hat U_b |\psi_1;\Delta t\rangle $.

However, the physical operation $\hat U_b$ defined above encompasses not only the action of the apparatus on the system being operated upon (i.e., the process $|\psi_2\rangle \to |\psi_2;\Delta t\rangle$) but also the transition from the state without the apparatus to the state with the apparatus (i.e., the process $|\psi_1\rangle \to |\psi_2\rangle$).
If we wish to make $|\psi_2\rangle$ more similar to $|\psi_1\rangle$, we may let $|\psi_1\rangle$ also include the same apparatus but with it switched off, i.e., it will not perform any operation. We denote this new state as $|\psi'_1\rangle$. By carrying out a similar derivation, we can likewise obtain a unitary operator $\hat U'_b$ supported in region $b$ such that $|\psi_2;\Delta t\rangle =\hat U'_b |\psi'_1;\Delta t\rangle $.
But even if $|\psi'_1\rangle$ includes the same apparatus, $\hat U'_b$ still incorporates the change of the apparatus from the “off” state to the “on” state.
If we wish the physical operation $\hat V_b$ to include only the action of the apparatus on the system being operated on (i.e., $|\psi_2;\Delta t\rangle =\hat V_b |\psi_2\rangle $), then we must require that the system being operated on, $|\psi_1\rangle$, be an eigenstate of the Hamiltonian $\hat H$. 
In other words, in the absence of the operating apparatus, the system $|\psi_1\rangle$ remains unchanged under time evolution (e.g., Witten chose $|\psi_1\rangle$ to be the vacuum state in Ref.~\cite{Witten2018}).
In this case, 
$|\psi_1;\Delta t\rangle \langle\psi_1;\Delta t| = |\psi_1\rangle \langle\psi_1|$,
and since, as stated earlier,
$\text{tr}_a\left(|\psi_1\rangle \langle\psi_1|\right) = \text{tr}_a\left(|\psi_2\rangle \langle\psi_2|\right) \text{ with } a\subset b$,
it follows that
$\text{tr}_b\left(|\psi_1;\Delta t\rangle \langle\psi_1;\Delta t|\right) = \text{tr}_b\left(|\psi_2\rangle \langle\psi_2|\right)$.
Combining this with the previously mentioned relation $\text{tr}_b\left(|\psi_1;\Delta t\rangle \langle\psi_1;\Delta t|\right) = \text{tr}_b\left(|\psi_2;\Delta t\rangle \langle\psi_2;\Delta t|\right)$, 
we obtain
$\text{tr}_b\left(|\psi_2;\Delta t\rangle \langle\psi_2;\Delta t|\right) = \text{tr}_b\left(|\psi_2\rangle \langle\psi_2|\right)$. 
This implies the existence of a unitary operator $\hat V_b$ supported in region $b$ such that
 $|\psi_2;\Delta t\rangle=e^{-i\hat H t}|\psi_2\rangle =\hat V_b |\psi_2\rangle $.

We have discussed three different meanings of ``physical operations" above, each corresponding to a local unitary operator $\hat U_b$, $\hat U'_b$, and $\hat V_b$.
However, it is worth noting that $|\psi_2;\Delta t\rangle$ and $|\psi_2\rangle$ include not only the system being operated on but also the apparatus (or the human) performing the operation.
This may differ from previous interpretations. Nevertheless, it is natural, since isolating the apparatus (or the human) from $|\psi_2;\Delta t\rangle$ generally leaves the operated system no longer in a pure state, owing to the interaction between the apparatus (or the human) and the system.
But for the sake of convenience in the following discussion, we will nevertheless treat the apparatus as independent of the quantum state $|\psi_2\rangle$.
Based on the previous derivation, we may rephrase Witten's statement as: ``If one asks not about mathematical operations in Hilbert space but about physical operations that are possible in the real world, then, since the apparatus performing the operation and the system being operated on are both made up of the same fundamental fields and obey causality, the only physical way that one can modify a quantum state brings about a unitary transformation."
Thus, even if we forcibly treat the apparatus as independent of the quantum state, causality directly leads us to the conclusion that a physical operation must be equivalent to a unitary operator supported in the region where the apparatus is located.
In this way, we bypass Witten’s ``perturbing the Hamiltonian (or refer to the third hypothesis in \cite{Eberhard1989})" and arrive at the final conclusion that physical operations are equivalent to unitary transformations.


\section{Schr\"odinger's cat paradox and A Possible Interpretation of Quantum Mechanics}
\label{inter}

In Section \ref{W}, complex macroscopic systems like apparatuses and humans can be described by quantum states and follow the Schr\"odinger equation, which implicitly assumes the completeness of quantum mechanics. Although the many-worlds interpretation \cite{EverettIII1957,Aharonov2008} aligns perfectly with the completeness of quantum mechanics, it requires people to accept the bizarre ``many-worlds scenario" that arises from Schr\"odinger's cat paradox.
This section will argue that in quantum field theory, we cannot derive Schr\"odinger's cat paradox as we do in non-relativistic quantum mechanics, which implies that we can get an interpretation satisfying the completeness of quantum mechanics without the need for the ``many-worlds scenario". For convenience, we will figuratively refer to this new interpretation as the ``one-world interpretation".
We also discuss the differences of this interpretation from traditional hidden-variable theories and its compatibility with relativity.

\subsection{
	Schr\"odinger's Cat Paradox in Quantum Field Theory}
\label{0}

In quantum mechanics, there is a fundamental assumption often cited as additional ``axiom (0)"\cite{Zurek2018,Zurek2009}: ``The states of composite quantum systems are represented by a vector in the tensor product of the Hilbert spaces of its components."  This leads to Schr\"odinger's cat paradox \cite{Aharonov2008a}.
We can refer to the mathematical symbols used in Ref. \cite{Aharonov2008} to demonstrate the derivation of the paradox.
let $D$ be a detector acting on the system $S$.
Let the states $|s_i\rangle$ be eigenstates of some observable and let the detector $D$ measure that observable. Let $|d_i\rangle$ be the state of the detector $D$ indicating that the system $S$ is in the state $|s_i\rangle$; 
 in other words, after measuring $S$ when it is in the state $|s_i\rangle$, the detector $D$ will be in the state $|d_i\rangle$.
Let the initial state of the detector be $|d_0\rangle$. Then, according to ``axiom (0)", the composite system of the system $S$ and the detector $D$ can be written as $|s_i\rangle|d_0\rangle$, and the measurement process can be expressed as
\begin{equation}
	\begin{split}
		\label{i0}
		|s_i\rangle|d_0\rangle\to|s_i,d_i\rangle\;.
	\end{split}
\end{equation}

In general, the initial state of the system S is a superposition $|s\rangle=\sum\limits_{i}c_i|s_i\rangle$, 
then unitary time evolution during the measurement takes this initial state to
\begin{equation}
	\begin{split}
		\label{p}
		|s\rangle|d_0\rangle=\sum\limits_{i}c_i|s_i\rangle|d_0\rangle\to\sum\limits_{i}c_i|s_i,d_i\rangle\;.
	\end{split}
\end{equation}
The final state is a superposition of different measurement outcomes, unable to yield a unique classical measurement result. In the Schr\"odinger's cat experiment, such a state is akin to being both dead and alive simultaneously. One might argue that the above derivation does not consider environmental factors; however, if we let the detector $D$ include the state of the environment, making the entire composite system isolated, the same paradox will still arise.
Various interpretations of quantum mechanics are presented in Ref. \cite{Aharonov2008} to resolve this paradox, including the well-known interpretation proposed by Wigner, where collapse occurs whenever a conscious human being observes a detector in a superposed state.
Here we specifically focus on the many-worlds interpretation, which holds that each measurement outcome $|s_i,d_i\rangle$ displayed in Eq. \eqref{p} is real, with different measurement outcomes corresponding to different worlds.

The above derivation of the cat paradox is based on non-relativistic quantum mechanics. However, in quantum field theory, the system $S$ and the detector $D$ are composed of the same fundamental fields. Therefore strictly speaking, we cannot fully distinguish between the system $S$ and the detector $D$.
Even if we consider quantum fields at different spatial points as different field operators, the interactions between fields at adjacent points lead to strong entanglement between different points in the quantum state.
Besides, the ubiquitous and constant interactions among fields cause field mixing, making it impossible to disentangle different types of fields \cite{Chen2024}. 		
Even the vacuum state, as fundamental and simple as it is, cannot be expressed as the direct product of ``vacuum states" of different regions, nor can it be expressed as the direct product of ``vacuum states" of different types of fields (bare vacuum).
Especially when the system $S$ and the detector $D$ are in the same spatial region during interactions, it becomes more challenging to distinguish them, rendering the use of axiom (0) inappropriate.
Therefore, in the framework of quantum field theory, the initial state of the ``composite" system formed by the detector $D$ and the system being measured $S$  cannot be written in the form of $|s\rangle|d_0\rangle$.

%
%

Although we cannot strictly distinguish the detector $D$ from the system being measured $S$ at the level of quantum states, we can distinguish them in space using reduced density matrices. The reduced density matrices characterize the information in a particular region. When the system $S$ and the detector $D$ do not overlap spatially, we can describe their respective states using reduced density matrices.
Let the quantum state $|s_1,d_0\rangle$ represent the system $S$ in state $s_1$ and the detector $D$ in state $d_0$, and let $|s_2,d_0\rangle$ represent the system $S$ in state $s_2$ and the detector $D$ in state $d_0$. Suppose the detector $D$ is located in region $A$ while the system $S$ is in region $a$. Since the detectors of both quantum states $|s_1,d_0\rangle$ and $|s_2,d_0\rangle$ are in state $d_0$, we have
\begin{equation}
	\begin{split}
		\text{tr}_a\left(|s_1,d_0\rangle \langle s_1,d_0|\right) =\text{tr}_a\left(|s_2,d_0\rangle \langle s_2,d_0|\right)\; .
	\end{split}
\end{equation}
According to the derivation in Section \ref{Gc}, it is known that there exists a  unitary operator $\hat U_a$ supported in region $a$, such that $|s_2,d_0\rangle=\hat U_a|s_1,d_0\rangle$.


Next, we consider a quantum state $|\psi_3\rangle\equiv c_1|s_1,d_0\rangle+c_2|s_2,d_0\rangle$.
Under time evolution, this state will develop into a superposition of different measurement outcomes, i.e., the Schrödinger-cat state $c_1|s_1,d_1\rangle+c_2|s_2,d_2\rangle$.
Assuming that $|s_1,d_0\rangle$ and $|s_2,d_0\rangle$ can be written as direct-product states, namely
$|s_1\rangle|d_0\rangle$ and $|s_2\rangle|d_0\rangle$, then
$|\psi_3\rangle=\big(c_1|s_1\rangle+c_2|s_2\rangle\big)|d_0\rangle$.
It is evident that
$\text{tr}_a\left(|\psi_3\rangle \langle \psi_3|\right)
=\text{tr}_a\left(|s_1,d_0\rangle \langle s_1,d_0|\right)$
and thus there exists a unitary operator $\hat V_a$ supported in region $a$ such that $|\psi_3\rangle=\hat V_a|s_1,d_0\rangle$.
According to Section \ref{W}, the unitary operator $\hat V_a$ supported in region $a$ constitutes a physical operation localized in region $a$.
Therefore, by manipulating the microscopic system located in region $a$, one can prepare the state $|\psi_3\rangle$ and thereby realize a Schrödinger-cat state.

However, when $|s_1,d_0\rangle$ and $|s_2,d_0\rangle$ cannot be written in direct-product form, in general there does not exist a unitary operator $\hat V_a$ supported in region $a$ such that $|\psi_3\rangle=\hat V_a|s_1,d_0\rangle$.
To illustrate the role played by entanglement in this context, we present a simple quantum-mechanical example.
Consider two systems, $a$ and $A$. The basis vectors describing system $a$ are $|\varphi_1\rangle_a,|\varphi_2\rangle_a,|\varphi_3\rangle_a$,
and the basis vectors describing system $A$ are $|\phi_1\rangle_A,|\phi_2\rangle_A,|\phi_3\rangle_A$.
We consider the following quantum state of the composite system consisting of 
$a$ and $A$:
\begin{equation}
	\begin{split}
		|s_1,d_0\rangle
		=b_1|\varphi_1\rangle_a|\phi_1\rangle_A
		+b_2|\varphi_2\rangle_a|\phi_2\rangle_A
		+b_3|\varphi_3\rangle_a|\phi_3\rangle_A\; .
	\end{split}
\end{equation}
The statement that system $A$ is in state $d_0$ means that system $A$ is described by the reduced density matrix
\begin{equation}
	\begin{split}
		\rho_A=\text{tr}_a\left(|s_1,d_0\rangle \langle s_1,d_0|\right)
		=b_1^*b_1|\phi_1\rangle_A \left._A\langle\phi_1|\right.
		+b_2^*b_2|\phi_2\rangle_A\left._A\langle\phi_2|\right.
		+b_3^*b_3|\phi_3\rangle_A\left._A\langle\phi_3|\right.\; .
	\end{split}
\end{equation}
Similarly, the statement that system $a$ is in state $s_1$ means that system $a$ is described by the reduced density matrix
$\rho_a
=b_1^*b_1|\varphi_1\rangle_a \left._a\langle\varphi_1|\right.
+b_2^*b_2|\varphi_2\rangle_a\left._a\langle\varphi_2|\right.
+b_3^*b_3|\varphi_3\rangle_a\left._a\langle\varphi_3|\right.$.
We define a unitary operator $\hat U_a$ acting on system $a$ such that
$$\hat U_a|\varphi_1\rangle_a=|\varphi_2\rangle_a,\qquad
\hat U_a|\varphi_2\rangle_a=|\varphi_3\rangle_a,\qquad
\hat U_a|\varphi_3\rangle_a=|\varphi_1\rangle_a \;.$$
We then define $|s_2,d_0\rangle$ as
\begin{equation}
	\begin{split}
		|s_2,d_0\rangle=\hat U_a|s_1,d_0\rangle
		=b_1|\varphi_2\rangle_a|\phi_1\rangle_A
		+b_2|\varphi_3\rangle_a|\phi_2\rangle_A
		+b_3|\varphi_1\rangle_a|\phi_3\rangle_A\; .
	\end{split}
\end{equation}
Thus, the statement that system $a$ is in state $s_2$ means that it is described by the reduced density matrix
$\rho'_a
=b_3^*b_3|\varphi_1\rangle_a \left._a\langle\varphi_1|\right.
+b_1^*b_1|\varphi_2\rangle_a\left._a\langle\varphi_2|\right.
+b_2^*b_2|\varphi_3\rangle_a\left._a\langle\varphi_3|\right.$.
Moreover, since the unitary operator $\hat U_a$ acts only on system $a$, the reduced density matrix of system $A$ remains unchanged, i.e.
$$
\rho'_A=\rho_A 
\quad \text{or equivalently} \quad
\text{tr}_a\left(|s_1,d_0\rangle \langle s_1,d_0|\right) =\text{tr}_a\left(|s_2,d_0\rangle \langle s_2,d_0|\right)\; .
$$
This shows that system $A$ corresponding to $|s_2,d_0\rangle$ indeed remains in state $d_0$.
Notice that an interesting feature of this example is that $|s_1,d_0\rangle$ and $|s_2,d_0\rangle$ are orthogonal, i.e. $\langle s_1,d_0|s_2,d_0\rangle=0$.
Combining the definitions of $|s_1,d_0\rangle$ and $|s_2,d_0\rangle$, we can write their superposition as
\begin{equation}
	\begin{split}
		|\psi_3\rangle
		&= c_1|s_1,d_0\rangle+c_2|s_2,d_0\rangle\\
		&
		=|\varphi_1\rangle_a
		\Big(c_1b_1|\phi_1\rangle_A+c_2b_3|\phi_3\rangle_A\Big)
		+|\varphi_2\rangle_a
		\Big(c_1b_2|\phi_2\rangle_A+c_2b_1|\phi_1\rangle_A\Big)
		+|\varphi_3\rangle_a
		\Big(c_1b_3|\phi_3\rangle_A+c_2b_2|\phi_2\rangle_A\Big)
		\; .
	\end{split}
\end{equation}
The reduced density matrix of system $A$ corresponding to $|\psi_3\rangle$ is
\begin{equation}
	\begin{split}
		\rho''_A
		&=b_1^*b_1|\phi_1\rangle_A \left._A\langle\phi_1|\right.
		+b_2^*b_2|\phi_2\rangle_A\left._A\langle\phi_2|\right.
		+b_3^*b_3|\phi_3\rangle_A\left._A\langle\phi_3|\right.\\
		&+c_2^*c_1 b_3^*b_1|\phi_1\rangle_A \left._A\langle\phi_3|\right.
		+c_1^* c_2 b_1^*b_3|\phi_3\rangle_A\left._A\langle\phi_1|\right.
		+c_1^* c_2 b_2^*b_1|\phi_1\rangle_A\left._A\langle\phi_2|\right.\\
		&+c_2^* c_1 b_1^*b_2|\phi_2\rangle_A \left._A\langle\phi_1|\right.
		+c_2^* c_1 b_2^*b_3|\phi_3\rangle_A\left._A\langle\phi_2|\right.
		+c_1^* c_2 b_3^*b_2|\phi_2\rangle_A\left._A\langle\phi_3|\right.
		\; .
	\end{split}
\end{equation}
Evidently, $\rho''_A\ne\rho_A$, i.e. $\text{tr}_a\left(|\psi_3\rangle \langle \psi_3|\right) \ne\text{tr}_a\left(|s_1,d_0\rangle \langle s_1,d_0|\right)$, which shows that there does not exist a unitary operator $\hat V_a$ acting only on system $a$ such that $|\psi_3\rangle=\hat V_a|s_1,d_0\rangle$.
This also demonstrates that it is impossible to obtain a Schrödinger-cat state solely by manipulating the measured microscopic system $S$ localized in region $a$.
\\

We can also give a fully quantum field–theoretic example to illustrate the differences between quantum field theory and quantum mechanics in the context of the Schrödinger’s cat problem.
We consider a special case in which $d_0$ in $|s_1,d_0\rangle$ and $|s_2,d_0\rangle$ is taken to be the vacuum, i.e.,
$$\text{tr}_a\left(|s_1,d_0\rangle \langle s_1,d_0|\right)
=\text{tr}_a\left(|s_2,d_0\rangle \langle s_2,d_0|\right)
=\text{tr}_a\left(|\Omega\rangle \langle \Omega|\right) \;,$$
meaning that the entire space contains only the measured microscopic system $S$ but no macroscopic detector $D$. Thus, we can directly abbreviate $|s_1,d_0\rangle$ and $|s_2,d_0\rangle$ as $|s_1\rangle$ and $|s_2\rangle$.
Since $|s_1\rangle$ and $|s_2\rangle$ describe microscopic systems, in nonrelativistic quantum mechanics we usually assume that we can prepare the superposition state $|s_1\rangle+|s_2\rangle$. However, in quantum field theory our operations must be represented by unitary local operators, and therefore even if $s$ is a microscopic system, we are not necessarily able to prepare the state $|s_1\rangle+|s_2\rangle$.
For the sake of concrete calculations, we take $|s_{1}\rangle$ and $|s_{2}\rangle$ to be coherent states in the free real scalar field theory.

We first briefly introduce some basic properties of coherent states, which have been proven in Ref. \cite{Chen2021}.
Let $|\phi\rangle$ be an eigenstate of the field operator $\hat\phi(\boldsymbol{x})$ with eigenvalue $\phi(\boldsymbol{x})$. The coherent state $|\phi_{\text{class}},\pi_{\text{class}}\rangle$ in the representation expanded by $|\phi\rangle$ is expressed as
\begin{equation}
	\begin{split}
		\label{phi state}	
		\langle\phi|\phi_{\text{class}},\pi_{\text{class}}\rangle
		=\mathcal{N'}
		&\exp\left\{
		-\frac{1}{2}\int d^3xd^3y \;
		\mathcal{E}(\boldsymbol{x},\boldsymbol{y})
		\left[\phi(\boldsymbol{x})-\phi_{\text{class}}(\boldsymbol{x})\right]
		\left[\phi(\boldsymbol{y})-\phi_{\text{class}}(\boldsymbol{y})\right]\right\}\\
		&\times \exp\left\{i\int d^3x\;\pi_{\textcolor{black}{\text{class}}}(\boldsymbol{x})\phi(\boldsymbol{x})
		\right\}
		\; ,
	\end{split}
\end{equation}
where $\mathcal{N'}$ is the normalization coefficient, and $\mathcal{E}(\boldsymbol{x},\boldsymbol{y})$ is defined by
\begin{equation}
	\label{E}
	\mathcal{E}(\boldsymbol{x},\boldsymbol{y})
	\equiv\int\frac{d^3p}{(2\pi)^3} e^{i\boldsymbol{p}\cdot(\boldsymbol{x}-\boldsymbol{y})} E_{\boldsymbol{p}}
	\; ,
	~~~~~~~~~~~~~~
	E_{\boldsymbol{p}}=\sqrt{\boldsymbol{p}^2+m^2}\;.
\end{equation}
The coherent state $|\phi_{\text{class}},\pi_{\text{class}}\rangle$ is determined by the corresponding classical fields $\phi_{\text{class}}$ and $\pi_{\text{class}}$ appearing in Eq. \eqref{phi state}.
Dividing the space into two regions, denoted as region $A$ and region $a$, and tracing out region $a$, we can obtain the reduced density matrix on region $A$ as
\begin{equation}\label{den1}
	\begin{split}
		\int\mathcal{D}\phi_a\;&\rho(\phi,\phi')
		\bigg|_{\phi(\boldsymbol{x}_a)=\phi'(\boldsymbol{x}_a),\;\boldsymbol{x}_a\in a}	
		=\int\mathcal{D}\phi_a\;
		\left<\phi|\phi_{\text{class}},\pi_{\text{class}}\right> 	\left<\phi_{\text{class}},\pi_{\text{class}}|\phi'\right>
		\bigg|_{\phi(\boldsymbol{x}_a)=\phi'(\boldsymbol{x}_a),\;\boldsymbol{x}_a\in a}	\\
		=|\mathcal{N}'|^2
		&\exp\left\{i\int_A d^3x\;\pi_{\text{class}}(\boldsymbol{x}_A)
		\Big[\phi(\boldsymbol{x}_A)-\phi'(\boldsymbol{x}_A)\Big]
		\right\}
		\\&
		\times\exp\left\{
		-\frac{1}{2}\int_A d^3x\int_A d^3y \;
		\mathcal{E}(\boldsymbol{x}_A,\boldsymbol{y}_A)
		\left[\phi(\boldsymbol{x}_A)-\phi_{\text{class}}(\boldsymbol{x}_A)\right]
		\left[\phi(\boldsymbol{y}_A)-\phi_{\text{class}}(\boldsymbol{y}_A)\right]\right\}
		\\&
		\times\exp\left\{
		-\frac{1}{2}\int_A d^3x\int_A d^3y \;
		\mathcal{E}(\boldsymbol{x}_A,\boldsymbol{y}_A)
		\left[\phi'(\boldsymbol{x}_A)-\phi_{\text{class}}(\boldsymbol{x}_A)\right]
		\left[\phi'(\boldsymbol{y}_A)-\phi_{\text{class}}(\boldsymbol{y}_A)\right]\right\}
		\\		
		\times\int\mathcal{D}&\varphi_a\;
		\exp\left\{-\int_a d^3x  \int_a d^3y\;
		\mathcal{E}(\boldsymbol{x}_a,\boldsymbol{y}_a)
		\varphi(\boldsymbol{x}_a)\varphi(\boldsymbol{y}_a)\right\}
		\\&
		\times\exp\left\{
		-\int_a d^3x  \int_A d^3y \;
		\mathcal{E}(\boldsymbol{x}_a ,\boldsymbol{y}_A)
		\varphi(\boldsymbol{x}_a)
		\left[\phi(\boldsymbol{y}_A)-\phi_{\text{class}}(\boldsymbol{y}_A)\right]\right\}
		\\&
		\times\exp\left\{
		-\int_a d^3x  \int_A d^3y \;
		\mathcal{E}(\boldsymbol{x}_a,\boldsymbol{y}_A)
		\varphi(\boldsymbol{x}_a)
		\left[\phi'(\boldsymbol{y}_A)-\phi_{\text{class}}(\boldsymbol{y}_A)\right]\right\}
		\;.
	\end{split}
\end{equation}
where $\boldsymbol{x}_A\in A$ and $\boldsymbol{x}_a\in a$, and $\int\mathcal{D}\phi_a$ denotes the functional integration over the field configurations restricted to region $a$ (i.e., over $\phi(\boldsymbol{x}_a)$ for all $\boldsymbol{x}_a\in a$).
Note that in Eq. \eqref{den1}, there are no appearances of $\phi_{\text{class}}(\boldsymbol{x}_a)$ and $\pi_{\text{class}}(\boldsymbol{x}_a)$. This indicates that Eq. \eqref{den1} only depends on the classical fields $\phi_{\text{class}}(\boldsymbol{x}_A)$ and $\pi_{\text{class}}(\boldsymbol{x}_A)$ in region $A$. Therefore, the reduced density matrix of a coherent state in a given region is completely determined by the classical fields in that region.
In other words, if two classical fields are identical in a certain region, 
then the reduced density matrices of their corresponding coherent states are also identical in that region.

We take $|s_1\rangle=|\phi_{\text{class}},\pi_{\text{class}}\rangle$ and $|s_2\rangle=|\phi'_{\text{class}},\pi'_{\text{class}}\rangle$, where the classical fields satisfy
$\phi_{\text{class}}(\boldsymbol{x}_A)= \pi_{\text{class}}(\boldsymbol{x}_A)=\phi'_{\text{class}}(\boldsymbol{x}_A)= \pi'_{\text{class}}(\boldsymbol{x}_A)=0$ for $\boldsymbol{x}_A\in A$.
Then, according to the properties of coherent states introduced above (noting that the vacuum state $|\Omega\rangle$ is a special coherent state with $\phi_{\text{class}}=\pi_{\text{class}}=0$), we have
$$\text{tr}_a\left(|s_1\rangle \langle s_1|\right)
=\text{tr}_a\left(|s_2\rangle \langle s_2|\right)
=\text{tr}_a\left(|\Omega\rangle \langle \Omega|\right) $$
This suggests that we can obtain $|s_1\rangle$ and $|s_2\rangle$ from the vacuum through unitary operators supported in region $a$.
At the same time, it implies that there exists a unitary operator $\hat U_a$ supported in region $a$ such that $|s_2\rangle=\hat U_a|s_1\rangle$.
Let $\psi_3=|s_1\rangle+|s_2\rangle$.
Since both $|s_1\rangle$ and $|s_2\rangle$ are coherent states, the reduced density matrix of $\psi_3$ can be directly written down. A straightforward calculation \cite{Chen2021} shows  that
$$\text{tr}_a\left(|\psi_3\rangle \langle \psi_3|\right)
\ne\text{tr}_a\left(|\Omega\rangle \langle \Omega|\right)
=\text{tr}_a\left(|s_1\rangle \langle s_1|\right) $$
This indicates that there does not exist a unitary operator $\hat V_a$ supported in region $a$ such that
$|\psi_3\rangle = \hat V_a |s_1\rangle.$
This shows that even if $S$ is a microscopic system localized in region $a$ and we are able to prepare its two states $|s_{1}\rangle$ and $|s_{2}\rangle$, we still cannot obtain their superposition $\psi_{3}=|s_{1}\rangle+|s_{2}\rangle$ by physical operations confined to region $a$ due to the entanglement between regions $a$ and $A$. This is markedly different from conventional nonrelativistic quantum mechanics.

Consequently, in quantum field theory, we have $|s,d_0\rangle\ne\sum\limits_{i}c_i|s_i,d_0\rangle$. As a result, even if we can write \eqref{i0} as $|s_i,d_0\rangle \to |s_i,d_i\rangle$, we cannot derive a formula similar to \eqref{p} (i.e., $|s,d_0\rangle=\sum\limits_{i}c_i|s_i,d_0\rangle\to\sum\limits_{i}c_i|s_i,d_i\rangle$).  Instead, we have
\begin{equation}
	\begin{split}
		\label{p'}
		|s,d_0\rangle\nrightarrow\sum\limits_{i}c_i|s_i,d_i\rangle\;.
	\end{split}
\end{equation}
This indicates that Schr\"odinger's cat paradox cannot be derived in quantum field theory in the same way as in nonrelativistic quantum mechanics.

The above discussion actually shows that the ability to express the initial state in a product form is crucial for deriving the Schrödinger’s cat paradox.
The traditional derivation of the Schrödinger’s cat paradox relies on two measurement-related ingredients:
(1) the measurement amplifies microscopic differences into macroscopic classical differences, i.e., $|s_i,d_0\rangle \to |s_i,d_i\rangle$;
(2) we can freely manipulate the microscopic measured system, that is, any unitary operator $\hat V_a$ supported in region $a$.
An initial state in  direct product form allows these two ingredients to be combined so that, one can obtain a superposition of vastly different macroscopic states by acting only on the microscopic system, thereby leading to the paradox.
In fact, in quantum field theory there also exist states of regions $a$ and $A$ that are not entangled (for example, eigenstates of field operators).
For such states, one can, in principle, use unitary operations  supported in region $a$ to derive the Schrödinger’s cat paradox (if the amplification effect still exists).
However, such non-entangled states are extremely nonclassical and idealized to the extent that they cannot naturally exist in the real world — even the vacuum state, the most fundamental state, is itself a highly entangled state.

%
It is worth noting that we do not deny the existence of superposition states of vastly different macroscopic states — for example, a quantum superposed cat that is both dead and alive at the same time.
We only contend that such a quantum superposed cat cannot be produced through a simple, traditional measurement process.
More specifically, although a state like $|\psi_3\rangle$, which could give rise to a paradox, may exist, it does not occur in actual measurement experiments. We do not deny that it might be possible to create such special states as $|\psi_3\rangle$ by using very complex and delicate nonlocal operations; however, that would constitute an entirely new type of experiment and would not contradict current simple measurement experiments.

We should also point out a few easily overlooked issues. The discussion above actually neglects the apparatus that prepares (or manipulates) the microscopic system $S$; a more rigorous treatment ought to include the apparatus as well.
Moreover, we acknowledge that we have not fully proven that the Schrödinger's cat paradox cannot arise in quantum field theory, but the foregoing discussion should be sufficient to underscore the importance of employing quantum field theory when addressing measurement issues and even questions of quantum interpretation.
Besides possessing intrinsic spatial entanglement, quantum field theory also imposes certain limitations on our ability to manipulate quantum systems. This constitutes a significant difference from nonrelativistic quantum mechanics.



%
%

\subsection{One-world Interpretation}
\label{owi}

To clarify the discussion, we need to provide a more detailed explanation of ``macroscopic state". A macroscopic state refers to a state described using classical concepts, such as a dead cat, a live cat, classical fields, a detector displaying measurement outcomes, and so on. Since classical descriptions are generally more ambiguous compared to quantum descriptions, a macroscopic state often corresponds to many quantum states. For example, many quantum states can describe ``a dead cat".
Even detectors displaying measurement outcomes correspond to multiple quantum states. This is because quantum states encompass not only specific displayed measurement readings but also factors like the average temperature of the detector, ubiquitous phonons in the detector, and so forth. More precisely, quantum states include the states of each atomic and molecular constituent composing the macroscopic detector.
Of course, there are also instances where each ``macroscopic state" corresponds to a single quantum state. For example, in free scalar field theory, the classical vacuum ($\phi=\pi=0$) corresponds uniquely to the vacuum state $|\Omega\rangle$ in quantum field theory.

As demonstrated in the previous subsection, there may be no Schr\"odinger's cat paradox in quantum field theory.
Therefore, a crucial possibility arises:
After the measurement process, the entire composite system evolves into a definite macroscopic state, rather than a superposition of macroscopic states. 
In other words, in the Schrödinger’s cat experiment, the cat would theoretically evolve into either being alive or being dead, rather than into a superposition of being both alive and dead.
We don't need (consciousness) to observe a detector to collapse it into a specific macroscopic state; instead, the detector automatically evolves to a definite macroscopic state.
This naturally leads to a new interpretation of quantum mechanics: quantum mechanics is complete, and the time evolution of quantum states can describe measurement processes, yielding a unique measurement outcome for each measurement.
This new interpretation and the many-worlds interpretation both consider quantum mechanics to be complete \cite{Aharonov2008}. 
However, in contrast to the many-worlds interpretation, this new interpretation argue that there aren't ``many worlds" but rather there is only ``one world".
Therefore, we may figuratively refer to this new interpretation as the ``one-world interpretation".
It is worth noting that this interpretation does not oppose the existence of superposition states of macroscopic states that are vastly different, such as a quantum superposed cat that is dead and alive at the same time.
However, this interpretation suggests that it is not possible to prepare a quantum superposed cat that is dead and alive by following the steps of the Schr\"odinger's cat experiment.

A natural follow-up question arises: where does the probability in quantum mechanics come from?
Looking back, the reason why probability was used to describe experimental results in the early days of quantum mechanics was that experiments with the same ``initial setup" yielded different results. Only by statistically analyzing the results of multiple experiments could patterns emerge.
In fact, however, the initial quantum state of each experiment is different.
The ``initial setup" includes the quantum state of the system being measured and the macroscopic state of the detector,
but a macroscopic state can correspond to multiple quantum states. Because we cannot control that every atom and molecule in the macroscopic detector remains unchanged, the initial quantum state of the entire system is not the same in each repeated experiment, which leads to the randomness of measurement results.
 Ref. \cite{Schonfeld2021} formulates a mechanism for how the first droplet in a cloud chamber track arises, which provides a concrete example illustrating that the randomness of detection outcomes originates from the initial conditions that include the quantum state of the detector itself.
The physics of a Geiger counter and a cloud chamber have a lot in common, then similar analysis can be applied to a Geiger counter as well \cite{Schonfeld2023}.

If we toss a coin multiple times and statistically analyze the results, we can obtain a stable probability distribution, from which we can extract some information about the coin. For example, if the probability of landing heads up is not $1/2$, it indicates that the coin may be asymmetric.
In quantum mechanics, measurements follow a similar principle. Through multiple repetitions of experiments and statistical analysis of the results, the randomness of the initial state can be averaged out, revealing information that remains constant across each experiment. Since the state of the system being measured remains the same in each experiment, the randomness inherent in the detector itself is averaged out, thereby revealing information about the system being measured.

Regarding the measurement process, a more specific but qualitative description is as follows. When the system being measured interacts with the detector, the composite system is in a highly unstable state. Subsequently, the entire system rapidly evolves into a stable macroscopic state, namely the state where the detector displays a specific reading.
If the system being measured is in an eigenstate of the observable before the measurement, then the stable macroscopic state of the composite system formed by the system and the detector is unique, and the composite system naturally evolves to a definite outcome.
If the system being measured is in a superposition of eigenstates of the observable before the measurement, then there are multiple stable state of the composite system.
Because the composite system is in a highly unstable state during the measurement process, even tiny changes in the initial state lead to the composite system evolving to different stable states, leading to various possible measurement outcomes. 
Therefore, the determination of which outcome the composite system evolves to depends on the precise initial quantum state of the composite system.

The mechanism for the formation of the first droplet in the Cloud Chamber, as presented in Ref. \cite{Schonfeld2021}, serves as an example illustrating the measurement process described above.
Specifically, when the decay product appear in the Cloud Chamber, the composite system is in a highly unstable state. At some time and in some location, a cluster of fortuitous size appears (as a result of thermal fuctuations) with an enormous ionization cross section, then the decay product wavefunction quickly becomes collimated at the location of the cluster, and the system rapidly evolves into a stable state—the state where the first droplet forms.

Finally, it is worth mentioning that in the one-world interpretation, there is no wave function collapse. Wave function collapse arises from treating the system under measurement as an independent entity from the detector at the quantum level, and describing the system being measured with a separate quantum state (wave function). For instance, in \eqref{p}, if we describe the system being measured as $|s\rangle$ and it evolves into $|s_i\rangle$ after the measurement process, we would say that the system has collapsed from $|s\rangle$ to $|s_i\rangle$.
However, in quantum field theory, since all matter in the world is composed of the same fundamental fields and there are interactions between the detector and the system being measured, the fields that make up the system being measured must also be part of the detector's composition according to the field mixing effect \cite{Chen2024}. 
Therefore, the system being measured cannot be described by a separate quantum state as in the non-relativistic case, especially when the measured system and the detector overlap in space and merge into a single entity during their interaction.
Consequently, there is no concept of wave function collapse in the one-world interpretation.

\subsection{Violation of Statistical Independence and  Compatibility with Relativity}
As demonstrated in the previous subsection, it can be seen that the one-world interpretation is actually a hidden-variable theory, where the hidden variables are the initial quantum states of the entire systems.
This naturally raises the question: Will this interpretation be ruled out by experiments \cite{Aspect1999} related to Bell inequalities?
Recalling the key formula used by Bell in deriving Bell's inequalities \cite{Bell1964}:
\begin{equation}
	\begin{split}
		P(\boldsymbol{a},\boldsymbol{b})
		=\int \mathrm d\lambda \; \rho(\lambda) 
		A(\boldsymbol{a},\lambda)  B(\boldsymbol{b},\lambda)
		\; ,
	\end{split}
\end{equation}
where the probability distribution $\rho(\lambda)$ of the hidden variable $\lambda$ is independent of the macroscopic states of the detectors, i.e. $\boldsymbol{a}$ and $\boldsymbol{b}$ (specifically $\boldsymbol{a}$ and $\boldsymbol{b}$ are polarizer settings).
In the one-world interpretation, the hidden variables $\lambda$ might superficially seem to correspond to the initial states of the detectors, however due to the failure of axiom (0), we cannot consider the two detectors separately, nor can we separate the detectors from the system being measured.
Therefore, the hidden variables are actually the initial quantum state of the entire system $|\psi\rangle$.
However, the quantum state $|\psi\rangle$ needs to satisfy certain constraints. In addition to ensuring that the initial states of the system being measured are consistent (i.e., their reduced density matrices of the system being measured are the same), it is also required that the corresponding macroscopic states of the detectors be $\boldsymbol{a}$ and $\boldsymbol{b}$, respectively.
This constraint indicates that the range of values of the hidden variable $\lambda$ is not the same for different macroscopic states $\boldsymbol{a}$ and $\boldsymbol{b}$ of the detectors. 
Consequently, the distribution function $\rho(\lambda)$ is actually dependent on the macroscopic states $\boldsymbol{a}$ and $\boldsymbol{b}$, and should be denoted as $\rho(\lambda|\boldsymbol{a},\boldsymbol{b})$ (More specifically, $\rho(\lambda|\boldsymbol{a},\boldsymbol{b})=\rho(|\psi\rangle| \text{the polarizer settings of $|\psi\rangle$ are $\boldsymbol{a}$ and $\boldsymbol{b})\ne \rho(\lambda)$}$), which differs from Bell's assumption.
In fact, the assumption that the hidden variables do not in any way depend on measurement settings, i.e. $\rho(\lambda|\boldsymbol{a},\boldsymbol{b})=\rho(\lambda)$, is commonly known as ``Statistical Independence" \cite{Hance2022,Hossenfelder2020}.
Therefore, although the one-world interpretation is a hidden variable theory, it violate Statistical Independence and does not lead to Bell's inequalities as in Ref. \cite{Bell1964}, and as a result, it will not be ruled out by experiments related to Bell's inequality.

In addition, the one-world interpretation does not belong to the category of ``nonlocal hidden-variable theory" as defined by Leggett in Ref. \cite{Leggett2003}, because the nonlocal hidden-variable theory defined by Leggett satisfies Statistical Independence.
Therefore, the one-world interpretation is not excluded by experiments \cite{Groeblacher2007} violating the inequality proposed by Leggett \cite{Leggett2003}.
In Section \ref{W}, it was proved that a physical operation must be equivalent to a local unitary operator. In the Sorkin-type impossible measurements problem, such a unitary operation is also referred to as a ``kick".
Studying the effect of a kick on measurements that are spacelike separated is central to the Fermi two-atom problem and the Sorkin-type impossible measurements problem.
In the Fermi two-atom problem, there is only one measurement, and there are no other measurements between the kick and this measurement. According to the derivation in Section \ref{lac}, it is evident that the kick cannot affect the measurement result if the measurement is space-like separated from the kick.

In the Sorkin-type impossible measurements problem, there are additional measurement processes between the kick and the spacelike separated measurement.
As mentioned in Section \ref{In}, traditional quantum mechanics is governed by two laws: the time evolution of quantum states and measurement theory. According to the traditionally accepted measurement theory, if there are additional measurement processes between the kick and the spacelike separated measurement, the kick has the ability to influence the results of the spacelike separated measurement, which violates relativistic causality.
However, within the framework of the one-world interpretation and in combination with the causality demonstrated in Section \ref{lac}, this troublesome Sorkin-type impossible measurements problem can be resolved.
The one-world interpretation holds that quantum mechanics is complete, and thus, the underlying physical laws behind the measurement theory corresponding to this interpretation are still the time evolution of quantum states. 
Therefore, regardless of whether there are additional measurements between the kick and the spacelike separated measurement, the entire system can always be described using the evolving quantum states over time, where ``state update" also falls under the time evolution of quantum states. 
Consequently, according to the causality demonstrated in Section \ref{lac}, the kick cannot influence measurements spacelike separated from it under any circumstances.
Furthermore in the one-world interpretation, each measurement process has a definite final outcome, and the quantum states encompass all the results of measurements, including those of intermediate measurements. Each measurement result can be read from the reduced density matrix of the region where the measurement occurs.
%
In this sense, the one-world interpretation is a deterministic theory, but surprisingly it neither violates causality nor undermines the completeness of quantum mechanics.



\section{Conclusion and Outlook}
\label{con}

We use the method of reduced density matrices to represent the local information of quantum states, thereby providing a general definition of causality in quantum field theory. We illustrate the time evolution of reduced density matrices in detail using the example of a free real scalar field theory, providing a very specific demonstration of causality. As for the proof of causality in general local quantum field theory, it is derived based on the mathematical conclusions of Ref. \cite{Eberhard1989} and further elaborated upon.
It should be emphasized that the third hypothesis in Ref. \cite{Eberhard1989} 
is precisely the statement about physical operations quoted from Witten in Section \ref{In}:
 ``the only physical way that one can modify a quantum state is by perturbing the Hamiltonian by which it evolves." However, the derivation in Section \ref{Gc} did not adopt this physical viewpoint, but only used the mathematical conclusions it leads to in Ref. \cite{Eberhard1989}.
It is only after we fully derive causality that we analyze Witten's statement about physical operations. Our analysis indicates that we can bypass ``perturbing the Hamiltonian" and arrive at the final conclusion that physical operations are equivalent to unitary transformations.

On the other hand, the intrinsic spatial entanglement of quantum field theory, together with the idea that all matter is composed of the same fundamental fields, indicates that the initial state of the measurement process cannot be written in a direct product form.
 Furthermore, even if two quantum states have identical reduced density matrices in a given region, their superposition may not retain the same reduced density matrix in that region due to the spatial entanglement inherent in quantum field theory.
 Schr\"odinger's cat paradox cannot be derived in quantum field theory in the same way as in nonrelativistic quantum mechanics, which may give rise to a new interpretation.  
This interpretation considers quantum mechanics to be complete (like the many-worlds interpretation) but does not need to introduce multiple worlds to deal with the cat paradox. Instead, the new interpretation proposes that there is only one world, which we figuratively refer to as the ``one-world interpretation".


In the one-world interpretation, measurement outcomes are deterministic, and they are determined by the initial quantum state of the entire system. However, we cannot maintain the exact same state of every atom and molecule in the detector (including the environment) in each repetition of the experiment, leading to the uncertainty of experimental results.
Clearly, this interpretation is a kind of hidden variable theory. However, as explained in Section \ref{inter}, because it violate Statistical Independence (for the meaning of ``Statistical Independence", see Ref. \cite{Hance2022, Hossenfelder2020}), it does not fall under Bell's local hidden variable theory \cite{Bell1964} or Leggett's definition of ``nonlocal hidden-variable theory" \cite{Leggett2003}. 
Therefore, it will not be ruled out by experiments related to Bell's inequality and Leggett's inequality.
In fact, the one-world interpretation closely resembles a superdeterministic theory \cite{Hossenfelder2020}; however, unlike traditional superdeterministic approaches, it maintains that nonlocal quantum mechanics is a complete theory.
Combining the causality defined in this paper, it can be concluded that the one-world interpretation is compatible with special relativity, and situations involving superluminal transmission of information do not arise, nor does the Sorkin-type impossible measurements problem.

%

While our work provides a framework to harmoniously integrate relativistic causality, quantum non-locality, and quantum measurement, we still lack a quantitative description of the one-world interpretation. Most crucially, we are uncertain about the correspondence between quantum states and macroscopic states, where ``macroscopic states" refer to states described using classical concepts, as introduced in Section \ref{owi}. For some simple cases, we can find correspondences between quantum states and macroscopic states \cite{Chen2021}.
However, it is often challenging to find the corresponding quantum state for general complex macroscopic states. For instance, it is difficult to determine the quantum state describing a cat.

In addition to understanding the correspondence between macroscopic states and quantum states, there are many fundamental questions about the measurement process that require further investigation. For example, how to write down the entire initial quantum state of a composite system, or how to quantitatively demonstrate that microscopic changes in the initial state can be amplified into macroscopic changes during the measurement process. Another question is how to quantitatively demonstrate that the final quantum state can evolve into a definite macroscopic state rather than a superposition state.
Perhaps a method worth considering is to use a bound state containing a large number of atoms to simulate the detector in quantum field theory.
However, while we aspire for the detector model to be entirely quantum mechanical, semi-classical detector models such as those in Ref. \cite{Schonfeld2021} and Ref. \cite{Schonfeld2023} can also address some of these questions and provide insights for further research. 

%

Finally, it should be clarified that we have not provided a rigorous proof that Schr\"odinger's cat paradox cannot arise within the framework of quantum field theory (QFT).
Therefore, the question of whether the cat paradox can arise in QFT merits further careful investigation.
However, because the initial state of the measurement process in QFT cannot be expressed as a direct product state in the same way as in nonrelativistic quantum mechanics, one cannot obtain the Schr\"odinger's cat paradox in QFT by the simple, conventional route that exploits the two measurement-related ingredients described in Section \ref{0}.
Ultimately, investigation of this issue inevitably requires confronting the aforementioned problems related to the development of the one-world interpretation.

\section{Acknowledgments}
We gratefully acknowledge fruitful conversations with Bartek Czech.
We also thank Qi Chen and Weijun Kong for helpful discussions.

\appendix

\section{The derivation of the free field propagator}
\label{Sss}

The propagator for a quantum harmonic oscillator, denoted as $\langle q|{\mathrm e}^{-i\hat H_o t}|q_1\rangle$, takes the form ${\mathrm e}^{iS(q,q_1;t)}$, where $S(q,q_1;t)$ is the extreme of the action under fixed path-boundaries (i.e., for fixed $q(0)=q$ and $q(t)=q_1$).
The free field theory described by the Hamiltonian $\hat H 
=\int \mathrm d^3x\;
\Big[\frac{1}{2}\hat\pi^2(\boldsymbol{x})
+\frac{1}{2}\Big(\nabla\hat\phi(\boldsymbol{x})\Big)^2
+\frac{1}{2}m^2\hat\phi^2(\boldsymbol{x})\Big]$
can be regarded as a collection of many harmonic oscillators, therefore we can similarly guess that the field propagator$\langle\phi|	{\mathrm e}^{-i\hat H t}|\phi_1\rangle$ also takes the same form:
\begin{equation}
	\begin{split}
		\label{ZeS}
		\langle\phi|	{\mathrm e}^{-i\hat H t}|\phi_1\rangle 
		&=N(t){\mathrm e}^{iS(\phi,\phi_1;t)}
		\; ,
	\end{split}
\end{equation}
where $S(\phi,\phi_1;t)$ is the extremum of the action evaluated for fixed initial state $\phi_1$ and final state $\phi$, and $N(t)$ is independent of $\phi$ and $\phi_1$. In fact, substituting the solutions satisfying the Euler-Lagrange equations into $S=\int \mathrm d^4x\; \mathcal{L}(\phi(x),\dot\phi(x))$ yields 
\begin{equation}\label{S}
	\begin{split}
		S(\phi,\phi_1;t)
		&=\frac{1}{2}\int \mathrm d^3x \mathrm d^3y\; G(\boldsymbol{x}-\boldsymbol{y};t)\Big[\phi(\boldsymbol{x})\phi(\boldsymbol{y})
		+\phi_1(\boldsymbol{x})\phi_1(\boldsymbol{y})\Big]
		\\&\quad	
		-\int \mathrm d^3x \mathrm d^3y\; g(\boldsymbol{x}-\boldsymbol{y};t)\phi(\boldsymbol{x})\phi_1(\boldsymbol{y})  
		\;,
	\end{split}
\end{equation}
where
\begin{equation}
	\begin{split}
		\label{Gag}
		G(\boldsymbol{x}-\boldsymbol{y};t)
		&=\int \frac{\mathrm d^3p}{(2\pi)^3} 	 \frac{p^0 \cos(p^0t)}{\sin(p^0t)}
		{\mathrm e}^{i\boldsymbol{p}\cdot(\boldsymbol{x}-\boldsymbol{y}) }\;,\\
		g(\boldsymbol{x}-\boldsymbol{y};t)
		&=\int \frac{\mathrm d^3p}{(2\pi)^3} 	\frac{p^0}{\sin(p^0t)}
		{\mathrm e}^{i\boldsymbol{p}\cdot(\boldsymbol{x}-\boldsymbol{y}) }\;.
	\end{split}
\end{equation}

Eq. \eqref{ZeS} is just our conjecture regarding the field propagator $\langle\phi|{\mathrm e}^{-i\hat H t}|\phi_1\rangle$. We need to verify that it reduces to $\prod\limits_x \delta(\phi(x)-\phi_1(x))$ at $t=0$, and satisfies the Schr\"odinger equation for any $t$:
\begin{equation}
	\begin{split}
		\label{xdefc}
		i\frac{\partial}{\partial t}\langle\phi|	{\mathrm e}^{-i\hat H t}|\phi_1\rangle   
		&=\hat H
		\langle\phi|	{\mathrm e}^{-i\hat H t}|\phi_1\rangle 	\;.
	\end{split}
\end{equation}
Note that in Eq. \eqref{xdefc}, the first $\hat H$ on the right-hand side actually represents the Hamiltonian in the representation defined by the eigenstates of the field operator $\hat\phi(x)$. Since the entire derivation involves only one representation, writing it this way does not introduce ambiguity. In the representation defined by the eigenstates of the field operator $\hat\phi(x)$, we have $\hat\phi(x)=\phi(x)$ and $\hat\pi(x)=-i\frac{\delta}{\delta \phi(x)}$.

To handle the Schr\"odinger equation \eqref{xdefc}, we need to compute $\int \mathrm d^3x\;\hat\pi^2(x)\langle\phi|	{\mathrm e}^{-i\hat H t}|\phi_1\rangle $.
Next, let's proceed to compute it.
Firstly, according to Eq. \eqref{Gag}, it is straightforward to calculate the specific expressions for $\frac{\partial}{\partial t}G(\boldsymbol{x}-\boldsymbol{y};t)$ and $\frac{\partial}{\partial t}g(\boldsymbol{x}-\boldsymbol{y};t)$:
\begin{equation}
	\begin{split}
		\label{dt1}
		\frac{\partial}{\partial t}G(\boldsymbol{x}-\boldsymbol{y};t)
		&=-\int \frac{\mathrm d^3p}{(2\pi)^3} 	 \frac{\left(p^0\right)^2}{\sin^2(p^0t)}
		{\mathrm e}^{i\boldsymbol{p}\cdot(\boldsymbol{x}-\boldsymbol{y}) }\; , \\
		\frac{\partial}{\partial t}g(\boldsymbol{x}-\boldsymbol{y};t)
		&=-\int \frac{\mathrm d^3p}{(2\pi)^3} 	\frac{\left(p^0\right)^2\cos(p^0t)}{\sin^2(p^0t)}
		{\mathrm e}^{i\boldsymbol{p}\cdot(\boldsymbol{x}-\boldsymbol{y}) }\;.
	\end{split}
\end{equation}
Secondly, utilizing Eq. \eqref{dt1}, we can obtain the following formulas:
\begin{equation}
	\begin{split}
		\label{GG}
		\int \mathrm d^3x\; G(\boldsymbol{x}-\boldsymbol{y};t)G(\boldsymbol{x}-\boldsymbol{z};t)&=-\frac{\partial}{\partial t}G(\boldsymbol{z}-\boldsymbol{y};t)
		+\nabla_z^2\delta^3(\boldsymbol{z}-\boldsymbol{y})
		-m^2\delta^3(\boldsymbol{z}-\boldsymbol{y}) ,\\
		\int \mathrm d^3x\; G(\boldsymbol{x}-\boldsymbol{y};t)g(\boldsymbol{x}-\boldsymbol{z};t)
		&=-\frac{\partial}{\partial t}g(\boldsymbol{z}-\boldsymbol{y};t)\; ,\\
		\int \mathrm d^3x\; g(\boldsymbol{x}-\boldsymbol{y};t)g(\boldsymbol{x}-\boldsymbol{z};t)
		&=-\frac{\partial}{\partial t}G(\boldsymbol{z}-\boldsymbol{y};t)\; .
	\end{split}
\end{equation}
Finally, utilizing \eqref{GG}, the expression for $\int \mathrm d^3x\;\hat\pi^2(x)\langle\phi|{\mathrm e}^{-i\hat H t}|\phi_1\rangle $ can be obtained as
\begin{equation}
	\begin{split}
		\label{pi^2}
		&\int \mathrm d^3x\;\hat\pi^2(x)\langle\phi|	{\mathrm e}^{-i\hat H t}|\phi_1\rangle \\
		&=-i \int \mathrm d^3x\; G(\boldsymbol{0};t)
		\langle\phi|	{\mathrm e}^{-i\hat H t}|\phi_1\rangle     \\
		&\quad+\Bigg[
		\int \mathrm d^3y \mathrm d^3z \;\phi(\boldsymbol{y})\phi(\boldsymbol{z})
		\int \mathrm d^3x \;G(\boldsymbol{x}-\boldsymbol{y};t)G(\boldsymbol{x}-\boldsymbol{z};t)\\
		&\quad\qquad+\int \mathrm d^3y \mathrm d^3z \;\phi_1(\boldsymbol{y})\phi_1(\boldsymbol{z})
		\int \mathrm d^3x \;g(\boldsymbol{x}-\boldsymbol{y};t)g(\boldsymbol{x}-\boldsymbol{z};t)\\
		&\quad\qquad-
		\int \mathrm d^3y \mathrm d^3z \;\phi_1(\boldsymbol{y}) \phi(\boldsymbol{z})
		\int \mathrm d^3x \;g(\boldsymbol{x}-\boldsymbol{y};t)G(\boldsymbol{x}-\boldsymbol{z};t)\\
		&\quad\qquad-	\int \mathrm d^3y \mathrm d^3z \;\phi(\boldsymbol{y}) \phi_1(\boldsymbol{z})
		\int \mathrm d^3x \;G(\boldsymbol{x}-\boldsymbol{y};t)g(\boldsymbol{x}-\boldsymbol{z};t)
		\Bigg]
		\langle\phi|	{\mathrm e}^{-i\hat H t}|\phi_1\rangle
		\\  	
		&=-i \int \mathrm d^3x \; G(\boldsymbol{0};t)
		\langle\phi|	{\mathrm e}^{-i\hat H t}|\phi_1\rangle     \\
		&\quad-\left[
		\int \mathrm d^3x  \;\left(\nabla\phi(x)\right)^2	+m^2\int \mathrm d^3x \;\phi^2(x)
		\right] \langle\phi|	{\mathrm e}^{-i\hat H t}|\phi_1\rangle  \\
		&\quad-\frac{\partial}{\partial t}
		\Bigg[
		-2\int \mathrm d^3x\mathrm d^3y\; g(\boldsymbol{x}-\boldsymbol{y};t)\phi(\boldsymbol{x})\phi_1(\boldsymbol{y})
		\\&\qquad\qquad+
		\int \mathrm d^3x\mathrm d^3y\; G(\boldsymbol{x}-\boldsymbol{y};t)\Big[\phi(\boldsymbol{x})\phi(\boldsymbol{y})+\phi_1(\boldsymbol{x})\phi_1(\boldsymbol{y})\Big]
		\Bigg]
		\langle\phi|	{\mathrm e}^{-i\hat H t}|\phi_1\rangle \;.		
	\end{split}
\end{equation}

With Eq. \eqref{pi^2}, we immediately know that the expression on the right-hand side of the Schr\"odinger equation \eqref{xdefc} is given by
\begin{equation}
	\begin{split}
		\label{xdefcr}
		\hat H &\langle\phi|	{\mathrm e}^{-i\hat H t}|\phi_1\rangle   \\
		=&-i\frac{1}{2} \int \mathrm d^3x \; G(\boldsymbol{0};t)
		\langle\phi|	{\mathrm e}^{-i\hat H t}|\phi_1\rangle     \\
		&-\frac{1}{2}\frac{\partial}{\partial t}\Bigg[-2\int \mathrm d^3x\mathrm d^3y\; g(\boldsymbol{x}-\boldsymbol{y};t)\phi(\boldsymbol{x})\phi_1(\boldsymbol{y})
		\\&\qquad\qquad+
		\int \mathrm d^3x\mathrm d^3y\; G(\boldsymbol{x}-\boldsymbol{y};t)\Big[\phi(\boldsymbol{x})\phi(\boldsymbol{y})+\phi_1(\boldsymbol{x})\phi_1(\boldsymbol{y})\Big]
		\Bigg]
		\langle\phi|	{\mathrm e}^{-i\hat H t}|\phi_1\rangle \;.
	\end{split}
\end{equation}
The left-hand side of the Schr\"odinger equation \eqref{xdefc} can be directly computed as
\begin{equation}
	\begin{split}
		\label{xdefcl}
		i\frac{\partial}{\partial t}&\langle\phi|	{\mathrm e}^{-i\hat H t}|\phi_1\rangle   \\
		=&i\frac{1}{N(t)}\frac{d N(t)}{d t}
		\langle\phi|	{\mathrm e}^{-i\hat H t}|\phi_1\rangle     \\
		&-\frac{1}{2}\frac{\partial}{\partial t}\Bigg[-2\int \mathrm d^3x\mathrm d^3y\; g(\boldsymbol{x}-\boldsymbol{y};t)\phi(\boldsymbol{x})\phi_1(\boldsymbol{y})
		\\&\qquad\qquad+
		\int \mathrm d^3x\mathrm d^3y\; G(\boldsymbol{x}-\boldsymbol{y};t)\Big[\phi(\boldsymbol{x})\phi(\boldsymbol{y})+\phi_1(\boldsymbol{x})\phi_1(\boldsymbol{y})\Big]
		\Bigg]
		\langle\phi|	{\mathrm e}^{-i\hat H t}|\phi_1\rangle \;.
	\end{split}
\end{equation}

Substituting the specific expressions from Eq. \eqref{xdefcr} and Eq. \eqref{xdefcl} into the Schr\"odinger equation \eqref{xdefc}, we obtain an equation involving $N(t)$:    
\begin{equation}
	\begin{split}
		\frac{d }{d t}\ln N(t)
		=-\frac{1}{2} \int \mathrm d^3x\;G(\boldsymbol{0};t)\;.
	\end{split}
\end{equation}
This is consistent with the earlier assumption that $N(t)$ is independent of $\phi$ and $\phi_1$, indicating that our previous conjecture Eq. \eqref{ZeS} indeed satisfies the Schr\"odinger equation, and the specific expression for $N(t)$ is   
\begin{equation}
	\begin{split}\label{Nt}
		N(t)=\mathcal N 
		{\mathrm e}^{-\frac{1}{2}\int \mathrm d^3x\int dt\;G(\boldsymbol{0};t)}\;.
	\end{split}
\end{equation}

However, merely demonstrating that \eqref{ZeS} satisfies the Schr\"odinger equation as a solution is not sufficient. We also need to examine the behavior of $\langle\phi|	{\mathrm e}^{-i\hat H t}|\phi_1\rangle$ as $t\to 0$. Utilizing \eqref{Nt} and \eqref{S}, we can express \eqref{ZeS} in a more specific form:\\\\
\begin{equation}
	\begin{split}
		\label{NeS}
		&\langle\phi|	{\mathrm e}^{-i\hat H t}|\phi_1\rangle\\ 
		&=\mathcal N 
		{\mathrm e}^{-\frac{1}{2}\int \mathrm d^3x\int dt\;G(\boldsymbol{0};t)}
		\exp\Bigg\{
		-i\int \mathrm d^3x\mathrm d^3y\; g(\boldsymbol{x}-\boldsymbol{y};t)\phi(\boldsymbol{x})\phi_1(\boldsymbol{y})
		\\&\qquad\qquad\qquad
		+\frac{i}{2}\int \mathrm d^3x\mathrm d^3y\; G(\boldsymbol{x}-\boldsymbol{y};t)\Big[\phi(\boldsymbol{x})\phi(\boldsymbol{y})+\phi_1(\boldsymbol{x})\phi_1(\boldsymbol{y})\Big]
		\Bigg\}\; .
	\end{split}
\end{equation}
As $t\to 0$, it is easy to obtain from \eqref{Gag} that $G(\boldsymbol{x}-\boldsymbol{y};t)\to\frac{1}{t}\delta^3(\boldsymbol{x}-\boldsymbol{y})$ and $g(\boldsymbol{x}-\boldsymbol{y};t)\to\frac{1}{t}\delta^3(\boldsymbol{x}-\boldsymbol{y})$. Consequently, the behavior of \eqref{NeS} as $t\to 0$ can be obtained:
\begin{equation}\label{t0}
	\begin{split}
		&	\lim\limits_{t\to0}\langle\phi|	{\mathrm e}^{-i\hat H t}|\phi_1\rangle \\
		&=\mathcal N \lim\limits_{t\to0}
		{\mathrm e}^{-\frac{1}{2}\int \mathrm d^3x \delta^3(\boldsymbol{0})\ln(t)}
		{\mathrm e}^{	\frac{i}{2t}\int \mathrm d^3x \left[\phi(\boldsymbol{x})-\phi_1(\boldsymbol{x})\right]^2}\;.
	\end{split}
\end{equation}
Note that after regularization, the term $\delta^3(\boldsymbol{0})$ effectively becomes $\frac{1}{\mathrm dx^3}$. Expanding the integral in the exponential of Eq. \eqref{t0} into a product of exponentials, we obtain:
\begin{equation}
	\begin{split}
		\lim\limits_{t\to0}\langle\phi|	{\mathrm e}^{-i\hat H t}|\phi_1\rangle 
		&=\mathcal N \lim\limits_{t\to0}
		\prod_x \frac{1}{\sqrt t}		
		{\mathrm e}^{	\frac{i (\mathrm dx)^3}{2t} \left[\phi(\boldsymbol{x})-\phi_1(\boldsymbol{x})\right]^2}
		=	\prod_x \delta(\phi(x)-\phi_1(x))\;,
	\end{split}
\end{equation}
where $\mathcal N\equiv\prod\limits_x\left(\frac{(\mathrm dx)^3}{2\pi i}\right)$. This implies that \eqref{NeS} not only serves as a solution to the Schr\"odinger equation, but also represents the inner product between field operator eigenstates as $t \to 0$, indicating that \eqref{NeS} and \eqref{ZeS} are the correct expressions for the field propagator.

\newpage

\bibliographystyle{JHEP}
\bibliography{causality}

\end{document}